\documentclass[oldversion]{aa}
\usepackage{graphicx}
\usepackage{txfonts}
\usepackage{natbib}
\usepackage{subfigure}
\usepackage{color}

\newcommand{\numsim}{166}

\newcommand{\mfe}{$\langle \rm{[Fe/H]}\rangle$}

\begin{document}


   \title{The Dynamical and Chemical Evolution of Dwarf Spheroidal Galaxies}

   \author{Y. Revaz\inst{1} 
          \and 
	  P. Jablonka\inst{1,2}
	  \and 
	  T. Sawala\inst{3}
	  \and
	  V. Hill\inst{4}
	  \and
	  B. Letarte\inst{5}
	  \and
	  M. Irwin\inst{6}
	  \and
	  G. Battaglia\inst{7}
	  \and
	  A. Helmi\inst{8}
	  \and
	  M. D. Shetrone\inst{9}	  	  	  
	  \and
	  E. Tolstoy\inst{10}
	  \and
	  K.A. Venn\inst{11}
}

  \offprints{Y. Revaz}


  \institute{Laboratoire d'Astrophysique, \'Ecole Polytechnique F\'ed\'erale de Lausanne (EPFL), 1290 Sauverny, Switzerland\\
        \and
        University of Geneva, Observatory, 1290 Sauverny, Switzerland ; on leave from CNRS, UMR 8111\\
	\and	
        Max-Planck-Institut f\"ur Astrophysik, Karl-Schwarzschild-Stra\ss e 1, 85748 Garching bei M\"unchen, Germany\\
	\and
	Observatoire de la C\^ote d'Azur, CNRS UMR 6202, BP 4229, 06304 Nice  Cedex 4, France 	\\	
	\and	
	California Institute of Technology, MC105-24, Pasadena, CA 91125, USA\\
	\and	
	Institute of Astronomy, Madingley Road, Cambridge CB03 0HA, UK\\
	\and	
	European Southern Observatory, Karl-Schwarzschild-Stra\ss e 1, 85748 Garching bei M\"unchen, Germany\\
	\and	
	Kapteyn Astronomical Institute, University of Groningen, P.O. Box 800, 9700 AV Groningen, Netherlands\\
				\and	
	McDonald Observatory, University of Texas, Fort Davis, TX 79734, USA\\							
	\and	
	Department of Physics \& Astronomy, University of Victoria, Elliott Building, 3800 Finnerty Road, Victoria, BC, V8P 5C2, Canada \\
}	

   \date{Received -- -- 20--/ Accepted -- -- 20--}

 
  \abstract
   {



We present a large sample of fully self-consistent hydrodynamical
Nbody/Tree-SPH simulations of isolated dwarf spheroidal galaxies
(dSphs).  It has enabled us to identify the key physical parameters
and mechanisms at the origin of the observed variety in the Local
Group dSph properties.  The initial total mass (gas + dark matter) of
these galaxies is the main driver of their evolution. Star formation
(SF) occurs in series of short bursts.  In massive systems, the very
short intervals between the SF peaks mimic a continuous star formation
rate, while less massive systems exhibit well separated SF bursts, as
identified observationally.  The delay between the SF events is
controlled by the gas cooling time dependence on galaxy mass.  The
observed global scaling relations, luminosity-mass and
luminosity-metallicity, are reproduced with low scatter.  We take
advantage of the unprecedentedly large sample size and data
homogeneity of the ESO Large Programme DART, and add to it a few
independent studies, to constrain the star formation history of five
Milky Way dSphs, Sextans, LeoII, Carina, Sculptor and Fornax. For the
first time, [Mg/Fe] vs [Fe/H] diagrams derived from high-resolution
spectroscopy of hundreds of individual stars are confronted with model
predictions.  We find that the diversity in dSph properties may well
result from intrinsic evolution.  We note, however, that the presence
of gas in the final state of our simulations, of the order of what is
observed in dwarf irregulars, calls for removal by external
processes. }


   \maketitle

%

\section{Introduction}


Understanding the  dominant  physical processes  at the origin  of the
dynamical and chemical properties of dwarf spheroidal galaxies (dSphs)
is challenging.  The  binding  energy of  the interstellar  medium  of
these  low  mass systems, at the  faint  end of  the galaxy luminosity
function, is weak. The injection of energy, due to violent explosions
of supernovae
\citep{dekel86,mori97,maclow99,murakami99,mori02,hensler04,ricotti05,kawata06},
or the cosmic UV background during reionization
\citep{efstathiou92,barkana99,bullock00,mayer06}  may leave  dSphs totally devoid of gas
and consequently quench their   star formation. In this  picture,  the
majority of  the dSphs are fossils  of the  reionization epoch and are
characterized by an old stellar population \citep{ricotti05}.

However, observations offer evidence  for  more complex star  formation
histories   and  reveal a   clear variety    of  dwarf    galactic  systems
\citep{mateo98,dolphin02}.   The  spread   in  stellar  chemical
abundances, and in particular the low [$\alpha$/Fe] values compared to
Galactic halo stars at equal metallicity, are hardly compatible with an
early termination of star formation at the epoch of reionization
\citep[e.g.,][examples taken in relation to the galaxies studied in
this                                                  work]{harbeck01,
shetrone98,shetrone01,shetrone03,tolstoy03,geisler05,koch08}.   Whilst
some dSphs  are indeed  consistent with  rather short  star  formation
episodes, such as Sextans \citep{lee03} or Sculptor
\citep{babusiaux05},  others are characterized  by  much more extended
periods, like Carina
\citep{smecker-hane96,hurley-keller98} or  Fornax \citep{coleman08}.

Long durations of the star formation were early advocated by purely
chemical evolution models constrained by the dwarf metallicity
distributions and color-magnitude diagrams \citep{ikuta02,
  lanfranchi04}. Subsequent simulations of dwarf galaxies introduced
the role of the dark matter coupled to the stellar feedback
\citep{ferrara00}, and later the full dynamical physics of the gas and
dark matter, by means of N-Body+SPH treatment.  Along this line,
\citet{marcolini06,marcolini08} concluded that a prolonged (compared
to instantaneous) star formation requires an external cause for gas
removal, which cannot be due to galactic winds.  Intermittent episodes
of star formation were at the focal point of the analysis by
\citet{stinson07}.  They naturally arose from the alternation between
feedback and cooling of the systems. \citet{valcke08} confirmed their
self-regulated form.  These authors also found a gradual shift of the
star formation towards the inner galactic regions.  \citet{kawata06}
had looked for evidence of spatial variation as well, in the form of
metallicity gradients, but had to stop their simulations at redshift
1.  Likewise, considering a cosmological box as initial conditions
instead of individual halos, \citet{read06} stopped their simulations
early, and focused on the smallest and most metal-poor dwarf galaxies.

In all these works, gas remains at the end of the dSph evolutions.
The resolution of this problem constitutes a challenge.
\citet{mayer06} performed simulations of gas-rich dwarf galaxy
satellites orbiting within a Milky Way-sized halo and studied the
combined effects of tides and ram pressure.  They showed that while
tidal stirring produces objects whose stellar structure and kinematics
resemble that of dSphs, ram-pressure stripping is needed to entirely
remove their gas.  \citet{salvadori08} proposed a semi-analytical
treatment in a hierarchical galaxy formation framework and achieved
the smallest final gas fraction.

Despite  real   limitations,    such    as  scarce   comparisons  with
observations, incomplete  time-evolution,    or  ad      hoc
parameterizations, we are witnessing a rapid convergence toward
understanding  the formation and evolution   of dSphs. A critical step
forward must be   undertaken with a large set   of simulations to   be
confronted with an equally  broad  sample of data.  In  particular, the
chemical imprints resulting  from different hypotheses  have  not
yet been fully capitalized on.

The VLT/FLAMES instrument, with fiber links to the GIRAFFE and UVES
spectrographs, has enabled a revolution in spectroscopic studies of
resolved stellar populations in nearby galaxies.  It is now possible
to measure the abundances of a wealth of chemical elements for more
than 100 stars at once.  Our ESO-Large Programme DART (Dwarf
Abundances and Radial velocity Team) is dedicated to the measure of
abundances and velocities for several hundred individual stars in a
sample of three nearby dSph galaxies: Sculptor, Fornax, and Sextans.
We have used the VLT/FLAMES facility in the low resolution mode to
obtain CaII triplet metallicity estimates, as well as accurate radial
velocities out to the galaxies' tidal radii \citep{tolstoy04,
  battaglia06, helmi06, battaglia08, battaglia08b}.  Each of the three
galaxies has also been observed at high resolution for about 80 stars
in their central regions, to obtain detailed abundances for a range of
interesting elements such as Mg, Ca, O, Ti, Na, Eu. \citep[Hill et al
  in preparation; Letarte et al.  in preparation]{venn05,letarte07}.

In the following, we take advantage of the statistically significant
DART sample and data homogeneity, and include some recent independent
studies, to constrain the star formation history of five Milky Way
dSphs, Sextans, LeoII, Carina, Sculptor, and Fornax.  For the first
time, populated [$\alpha$/Fe] vs [Fe/H] diagrams can be confronted
with model predictions. Our first goal is to establish how well one
can reproduce the apparent diversity of dSph star formation histories
in a common scheme.  We choose to model galaxies in isolation, as this
is the only way to control the effect of all parameters at play, and
to understand the dominant physical processes.  We will try to see if
a complex star formation history may result from intrinsic evolution
or if external processes are necessary.  We have performed an
unprecedentedly large number of simulations. Not only do they account
for the gravity of the dark matter and baryons, but they also contain
a large number of additional physical mechanisms: metal-dependent gas
cooling above and below $10^4\,\rm{K}$, star formation, SNIa and SNII
energy feedback and chemical evolution.  We focus on the luminosity,
star formation history and metallicity properties of dSphs, rather
than on their dynamical properties, which turn out to be less
constraining.

The paper is organized as follows: The code and the implementation of
physical processes are described in Section~\ref{code}.  The initial
conditions are detailed in Section~\ref{model}.  The presentation of
the results is split in three parts: Section~\ref{models} focuses
on the global evolution of the galaxies and discusses the main driving
parameters, while Section~\ref{global_relations} is devoted to the
scaling relations. The detailed analysis of the chemical properties of
Sextans, Leo II, Carina, Sculptor, and Fornax are treated in
Section~\ref{generic_models}. Section ~\ref{discussion} offers a
physical interpretation of the results. Section~\ref{summary}
summarizes our work.



                                                         


\section{The code}\label{code}


We have adapted the code \emph{treeAsph} originally developed by
\citet{serna96}, with further developments presented in
\citet{alimi03}.  The chemical evolution was introduced by
\citet[Poirier, PhD thesis]{poirier03} and \citet{poirier02}.  For the
sake of simplicity, we recall below the main features and the general
philosophy of the algorithms.



\subsection{Dynamics}

All gravitational forces are computed under the \emph{tree} algorithm
proposed by \citet{barnes86} \citep[see also][]{hernquist87}.  This
technique is based on a hierarchical subdivision of space into cubic
cells.  One approximates the forces due to a cluster of particles
contained in a cubic cell and acting on a particle $i$ by a
quadrupolar expansion of the cluster gravitational potential.  This is
done under the condition that the size of the cubic cell is small
compared to its distance to the particle $i$.  The ratio between size
and distance must be smaller than a tolerance parameter, $\theta$,
fixed to 0.7 in our simulations \citep{hernquist87}.  Indeed, under
this condition, the internal distribution of the particles within
cells can be neglected. Consequently, the number of operations needed
to compute the gravitational forces between $N$ particles scales as
$\sim N\log N$, instead of $\sim N^2$ if one were to consider each
individual pair of particles.


The hydrodynamics of the gas is followed in the Lagrangian Smooth
Particle Hydrodynamics (SPH) scheme.  Its allows to describe an
arbitrarily shaped continuous medium with a finite number of particles
\citep{lucy77,gingold77}, see \citet{monaghan92} and \citet{price05}
for reviews.  Each gas particle has its mass spatially smeared out by
a smoothing kernel $W$ (here a spline function).

Unlike the gravitational forces, which are determined from the
interactions with all other particles in the system, the
hydrodynamical forces result from the contributions of a modest number
of neighbors. The spatial resolution is determined by the smoothing
length $h_i$ associated with the particle $i$, computed through the
requirement that a sphere of radius $h_i$ centered on particle $i$
contains $32$ neighbors.

The integration scheme is the symplectic leapfrog used with 
adaptative time-steps.

\subsection{Cooling}

At temperatures lower than $10^4\,\rm{K}$, cooling in primordial gas
is dominated by the molecule $H_2$.  \citet{galli98} have shown that
the cooling efficiency of $H_2$ is determined by its mass fraction
$\chi_{H_2}$.  Unfortunately, an accurate computation of $\chi_{H_2}$
is difficult and requires to take into account the complex processes
of the formation and destruction of $H_2$.  Following \citet{maio07},
we instead fix $\chi_{H_2} $to $10^{-5}$.  Once the gas is enriched
with metals, these are important to the cooling properties.  We
consider oxygen, carbon, silicon and iron \citep{maio07}, since they
are the most-abundant heavy atoms released during stellar evolution,
particularly by the SNe II and SNe Ia, that we follow in our
simulations.  We set the density of the free electrons over that of
hydrogen , ($n_{\rm{e}^-}/n_{\rm{H}}$) to $10^{-4}\,\rm{cm^{-3}}$.

\begin{figure}
  \resizebox{\hsize}{!}{\includegraphics[angle=0]{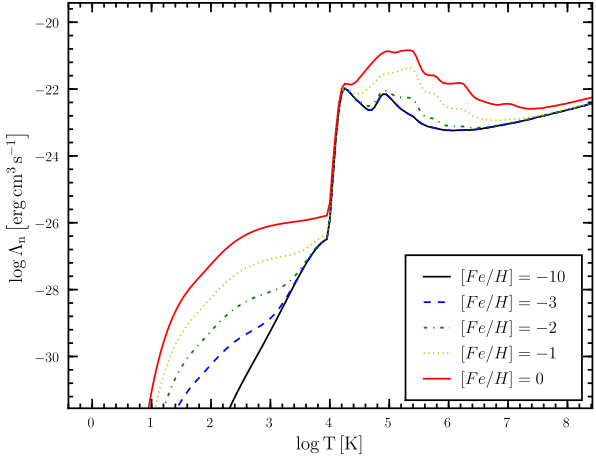}}
  \caption{Normalized cooling function as a function of temperature and metallicity.}
  \label{cooling_fct}
  \end{figure}

Above $10^4\,\rm{K}$, the cooling function is calculated following the
metallicity dependent prescription of \citet{sutherland93}.  The full
normalized cooling function is shown in Figure~\ref{cooling_fct}, for
a large range of metallicities.

A more detailed modeling of the gas cooling is not necessary, as long
as simulations are limited in spatial resolution.  Indeed, the cooling
of the gas is directly dependent on its density. Therefore, a limited
resolution smoothes the density fluctuations of the interstellar
medium. 


\subsection{Chemical evolution and stellar feedback}



The chemical enrichment of the interstellar medium (ISM) depends on
the interplay between different physical processes.  It requires us to
follow the rate at which stars form, the amount of newly synthesized
chemical elements, the mass and energy released during the different
stellar phases, and, finally, the mixing of the metal-enriched stellar
outflows with the ISM.  The computation of this cycle is done by
implementing the original equations of chemical evolution formalized
by \citet{tinsley80}, as closely as possible.

\subsection{Star formation}

We adopt the now classical recipe of \citet{katz92} and
\citet{katz96}.  A gas particle becomes eligible for star formation if
it is {\it i}) collapsing (negative velocity divergence) and {\it ii})
its density is higher than a threshold of $\rho_{\rm{sfr}}=1.67\times
10^{-25}\rm{g/cm^{3}}$.  However, we do not require the dynamical time
to be shorter than the sound crossing time (Jeans instability).


The gas particles, which satisfy  the above criteria, form stars
at a rate expressed by:
	\begin{equation}
	\frac{d \rho_\star}{dt} = \frac{c_\star}{t_{\rm{g}}}\rho_{\rm{g}},
	\label{sfr}
	\end{equation}
which mimics a Schmidt law \citep{schmidt59}. $c_\star$ is the
dimensionless star formation parameter, $t_{\rm{g}}$ is taken as the
maximum of the local cooling time and the free-fall time.

For a given time interval $\Delta t$, a gas particle of mass
$m_{\rm{g}}$ has a probability $p_{\star}$ to form a stellar particle
of mass $m_{\star}$, where $p_{\star}$ is defined by:
	\begin{equation}
	p_{\star} = \frac{m_{\rm{g}}}{m_{\star}}\left[ 1-\exp\left(  -\frac{c_\star}{t_{\rm{g}}}\Delta t  \right) \right].
	\label{pstar}
	\end{equation}
The new stellar particle is initially assigned the position and
velocity of its gas progenitor. Subsequently, gas and stellar
velocities are modified in order to conserve both energy and momentum.

Each stellar particle represent a cluster of stars, sharing the same
age and metallicity, whose initial mass function (IMF) $\Phi(m)$ is
described by a Salpeter law \citep{salpeter55}:
	\begin{equation}
	\Phi(m) = \left[ \frac{x+1}{ m_u^{x+1} - m_l^{x+1}} \right] m^x,
	\end{equation}
with $x=-1.35$,   $m_l=0.05\,\rm{M_\odot}$          and
$m_u=50\,\rm{M_\odot}$.


\subsection{Ejecta}

We neglect stellar winds, since they contribute little to the
evolution of the chemical elements that we consider (magnesium and
iron), and because the injection power to the ISM is dominated by SN
explosions \citep{leitherer92}.

The  amount of energy, mass  and metals ejected  by a stellar particle
during a time interval  $\Delta t$ is calculated  by considering the mass
of stars  exploding between $t$ and  $t+\Delta t$.  The  dependency of
the stellar lifetimes on  metallicity is taken into  account following
\citet[private communication]{kodama97}.


Hence, the feedback energy released by a stellar  particle in the time
interval $[t,t+\Delta t]$ is:
	\begin{equation}
	\Delta E_{\rm{SN}} = m_{\star,0}\,\,\left[ n_{\rm{II}}(t)\,E_{\rm{II}} + n_{\rm{Ia}}(t)\,E_{\rm{Ia}}  \right],
	\end{equation}
where $m_{\star,0}$ is the initial mass of the stellar particle, and
$n_{\rm{II}}(t)$ and $n_{\rm{Ia}}(t)$ are the corresponding number of
supernovae SNe II and SNe Ia per unit mass during $\Delta t$.  The
energy released by both SNe II ($E_{\rm{II}}$), and SNe Ia
($E_{\rm{Ia}}$), is set to $10^{51}\,\rm{erg}$.

With             $m_{\rm{II},l}=8\,\rm{M_{\odot}}$                 and
$m_{\rm{II},u}=50\,\rm{M_{\odot}}$ being the lowest and highest masses
of  stars exploding as  SNe II, and $m(t)$  being the mass of stars with
lifetime $t$, we can express $n_{\rm{II}}(t)$ as:
	\begin{equation}
	n_{\rm{II}}(t) = \int_{\max[m(t+\Delta t),m_{\rm{II},l}]}^{\min[m(t),m_{\rm{II},u}]}\frac{\Phi(m)}{m}\,dm.
	\end{equation}
To calculate $n_{\rm{Ia}}(t)$, we adopt the model of
\citet{kobayashi00} in which the progenitors of SNe Ia have
main-sequence masses in the range of $M_{\rm{p,l}}=3\,\rm{M_\odot}$ to
$M_{\rm{p,u}}=8\,\rm{M_\odot}$, and evolve into C+O white dwarfs
(WDs). These white dwarfs can form two different types of binary
systems (here labeled $i$), either with main sequence stars or with
red giants. Hence:
	\begin{equation}      
	n_{\rm{Ia}}(t)   =	\left(\int_{M_{\rm{p,l}}}^{M_{\rm{p,u}}}\frac{\Phi(m)}{m}\,dm\right)
			\sum_{i=1}^{i=2} b_i \int_{m_{i,1}}^{m_{i,2}} \frac{\Phi_d(m)}{m}\,dm,
	\end{equation}
with  $\Phi_d(m)$  the  distribution mass   function  of the companion
stars, $m_{i,1}=\max[m(t+\Delta t),M_{i,\rm{d,l}}]$,
$m_{i,2}=\min[m(t),M_{i,\rm{d,u}}]$.  The lifetime  of SNe Ia is $\sim
2-20$ and  $0.5-1.5$  Gyr, depending   on   the  binary system.    See
\citet{kobayashi00}  for  the lowest $M_{i,\rm{d,l}}$  and higher mass
$M_{i,\rm{d,u}}$ of the WDs progenitors  and  for the values of  $b_i$
which  weights  the probability  of  having one  or the  other  binary
system.

The supernova feedback energy is released in the form of thermal energy only at the
end of each dynamical time-step. This procedure avoids the thermal energy
to be dissipated instantaneously by the strong cooling above $10^4\,\rm{K}$,
and mimics the blast wave shocks of supernovae \citep{stinson06}.

 
The ejected gas mass fraction due to SNe Ia is given by:
	\begin{equation}
	\Delta m_{\rm{Ia}}(t) = m_{\rm{WD}} \, n_{\rm{Ia}}(t).
	\end{equation}
with $m_{\rm{WD}}=1.38\,\rm{M_\odot}$ being the mass of white dwarf.

%
%
%

The mass of each chemical element $k$ ejected by a stellar particle  is:
	\begin{equation}
	\Delta M_k = m_{\star,0}\,\,\left[ \Delta m_{k,\rm{II}}(t) + \Delta m_{k,\rm{Ia}}(t)  \right],
	\end{equation}
where:
	\begin{eqnarray}
	\Delta m_{k,\rm{II}}(t) & = &  \int^{m(t)}_{\max[m(t+\Delta t),m_{\rm{II},l}]} p_{k,\rm{II}}(m) \, \Phi(m)\,dm  \\\nonumber
	                   & + & z_{k}  \int^{m(t)}_{\max[m(t+\Delta t),m_{\rm{II},l}]} \left( 1- \omega(m) -  p_{k,\rm{II}}(m) \right)\, \Phi(m)\,dm  
	\end{eqnarray}
and
	\begin{equation}
	\Delta m_{k,\rm{Ia}}(t) = \Delta m_{\rm{Ia}}(t)\,p_{k,\rm{Ia}}(m). 
	\end{equation}
$\omega(m)$  is the remnant mass  fraction, which is the mass fraction
of a  black  hole, neutron  star or a  white  dwarf, depending  on the
initial   mass   $m$.    Values    of   $\omega(m)$  are     taken  from
\citep{kobayashi00}. $z_{k}$ is the original stellar abundance of the element $k$,
$p_{k,\rm{II}}(m)$ and $p_{k,\rm{Ia}}(m)$ are the stellar yields, i.e. the mass
fractions of newly produced and ejected  element $k$, coming from SNe II
and SNe I, respectively.  They are taken from \citet{tsujimoto95}.

Since the stellar particles   correspond to star  clusters, we  use the
     single   stellar    population mass-to-light ratios of
\citet{maraston98,maraston05}  to   calculate their   luminosities  in
$V$-band.  The effects of metallicity and age are taken into account.




\section{Initial Conditions}\label{model}


\subsection{Mass distribution}

We consider  dSphs in isolation. Gas  and dark  matter are
initially represented by pseudo-isothermal spheres:
	\begin{equation}
	\rho(r)=\frac{\rho_{\rm{c}}}{1+\left( \frac{r}{r_{\rm{c}}} \right)^2},
	\end{equation}
where $r$ is the radius, $r_c$ is the scale length of the mass
distribution, and $\rho_c$ the central mass density. The models are
truncated at $r_{\rm{max}}$.  The halo and gas mass distributions
differ by their central density, $\rho_{\rm{c,halo}}$ and
$\rho_{\rm{c,gas}}$, respectively.

As the the total mass inside a radius $r_{\rm{max}}$ is linearly
dependent on the central density, there is a proportionality relation
between the fraction of baryonic matter, $f_{\rm{b}}$, and the central
densities:
	\begin{equation}
	f_{\rm{b}} = \frac{M_{\rm{gas}}}{M_{\rm{gas}}+M_{\rm{halo}}} =
	\frac{\rho_{\rm{c,gas}}}{\rho_{\rm{c,gas}}+\rho_{\rm{c,halo}}}.
	\end{equation}
In    the  following,     we   will    use    $\rho_{\rm{c,tot}}$    =
$\rho_{\rm{c,gas}}+\rho_{\rm{c,halo}}$.    Using   $f_{\rm{b}}$    and
$\rho_{\rm{c,tot}}$, one can simply write:
	\begin{eqnarray}
	\rho_{\rm{c,gas}}  &=&  f_{\rm{b}}\,	\rho_{\rm{c,tot}}	   ,\nonumber\\
	\rho_{\rm{c,halo}} &=&  (1-f_{\rm{b}})\, \rho_{\rm{c,tot}}	   .
	\end{eqnarray}

Similarly to  the gas, the dark matter  halo evolves under the laws of
gravity.

We consider a  core  in the  initial dark  model profile. Whilst
cosmological simulations predict the formation of cuspy dark halos
\citep[and the references therein]{navarro97b,fukushige97,moore98,springel08}, 
our choice  is motivated by observational  evidences found  in normal,
low brightness and dwarf galaxies
\citep{blaisouellette01,deblok02,swaters03,gentile04,gentile05,
spekkens05,deblock05,deblok08,spano08}.   Measuring the inner slope of
the dSph profiles  is  very challenging,  nevertheless, \citet{battaglia08}
show that  the observed velocity dispersion  profiles of the Sculptor dSph
are best fitted by a cored dark matter halo.

\subsection{Velocities and temperature}

The   initial   velocities are   obtained    by assuming 
equilibrium, free of any  rotation.  For a spherical distribution,  we
can assume  that  the  velocity  dispersion  is  isotropic. It  can be
derived from the the second moment of the Jeans equation
\citep{binney87,hernquist93}. In spherical coordinates, one writes:
\begin{equation}
         \sigma^2(r) = \frac{1}{\rho(r)}\int_r^\infty\! dr' \,\rho(r')\, \partial_{r'} \Phi(r').
        \label{sr_sph}
\end{equation}
The halo velocities are randomly generated in order to fit the
velocity dispersion $\sigma^2$ at any given radius.  

The temperature $T$ of the gas is deduced from the virial equation:
\begin{equation}
         \lim_{r \rightarrow \infty} \frac{G M(r)}{r} = \frac{3\,k_{\rm{B}}\,T}{\mu m_{\rm{h}}},
        \label{u_gas}
\end{equation}
leading, for a pseudo-isothermal sphere, to:
\begin{equation}
        T = \frac{4\pi}{3} \frac{\mu m_{\rm{h}}}{k_{\rm{B}}}G\,\rho_{\rm{c,tot}} r_{\rm{c}}^2,
        \label{T_gas}
\end{equation}
where $k_{\rm{B}}$ is the Boltzmann constant, $m_{\rm{h}}$ the proton mass and $\mu$ 
the mean  molecular weight of the gas.


\subsection{Initial Parameters}\label{parameters}

All simulations  start with   an  initial radius of   $r_{\rm{max}}  =
8\,\rm{kpc}$, distance at which the gas and  dark matter densities are
about $1/1000$  of the central ones.  We  consider two  different core
radii, $r_{\rm{c}}$, of 0.5  and  1 kpc.  The   choice of the  central
total  density  $\rho_{\rm{c,tot}}$   (dark matter   +  gas)  uniquely
determines the initial total mass of  the system, $M_i$, which we vary
over  a range of $2\times 10^8$  to  $9\times 10^8\,\rm{M_\odot}$.  We
investigate   the  effect of   the initial    baryonic  mass fraction,
$f_{\rm{b}}$, by  varying it from 0.1 to  0.2.   Indeed, this helps in
disentangling the influence of the  total gravitational potential from
that of the gas mass.  The masses of the gas and halo particles remain
constant  at      $1.4\times 10^4\,\rm{M_\odot}$     and    $9.3\times
10^4\,\rm{M_\odot}$, respectively.  The    corresponding gravitational
softening lengths are  $0.1$ and $0.25\,\rm{kpc}$.   As a consequence,
the  simulations  start  with   $4000$  to  $20'000$   particles.  The
variation in  number of particles  is  therefore at  most a  factor 5,
hence a factor  of  1.7 in spatial   resolution, which justifies   the
choice of fixed   softening  lengths.  The star    formation parameter
$c_\star$ is varied  from $0.01$ to  $0.3$.  The  initial mass of  the
stellar  particles  is  $980\,M_{\odot}$, corresponding to   about one
tenth of the initial mass of the gas particles.


\section{Models}\label{models}


We performed~\numsim\, simulations   to  understand the role    of each
parameter at   play, and   to  identify  a  series of  generic  models
reproducing the observations.  The complete list of the simulations is
given  in  Tab.~\ref{parameters_1}, \ref{parameters_2}    and    \ref{parameters_3}     of
Appendix~\ref{appendix1}.   The      models   have been       run  for
$14\,\rm{Gyr}$.

\subsection{Modes}

As  mentioned in Section~\ref{model}, the  initial sphere of DM+gas is
in  equilibrium under   adiabatic conditions.   At  the  onset of  the
simulations, the energy loss due to cooling causes the gas to sinks in
the potential  well and contract.   The total potential is  deepen not
only  due to  the  central  increase in  gas  density, but  also  as a
consequence  of the halo  adiabatic  contraction.   Despite the  large
increase in density,  the gas temperature  is kept nearly constant due
to the strong hydrogen recombination cooling above $10^4\,\rm{K}$ (see
Fig.~\ref{cooling_fct}).     Therefore,   the    gravitational  energy
recovered  from  the deepening  of  the potential  well  is dissipated
nearly instantaneously. For   densities above $\rho_{\rm{sfr}}$,   the
evolution of  the model depends on  supernova heating, directly linked
to the star formation rate.

\begin{figure}
\resizebox{\hsize}{!}{\includegraphics[angle=0]{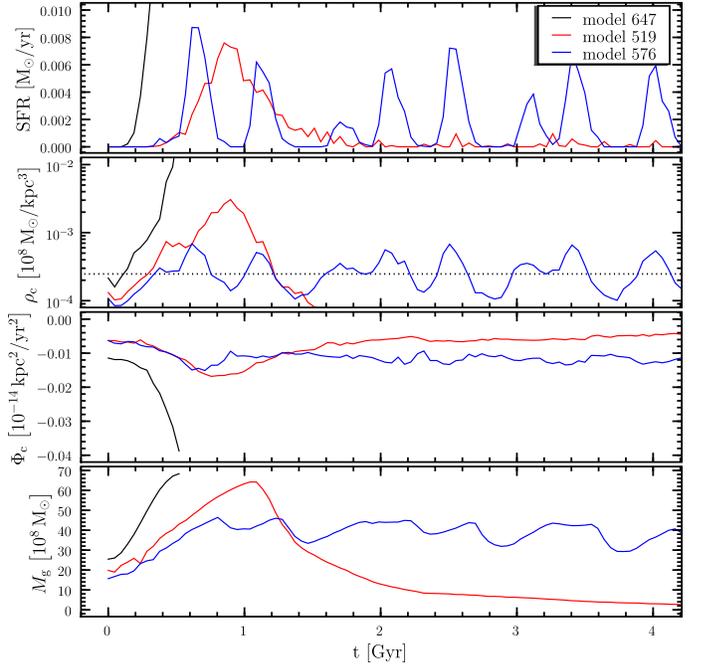}}
\caption{Time evolution  of the star  formation rate,  $\rm{SFR}$, the
gas central density, $\rho_{\rm{c}}$, the galaxy central gravitational
potential, $\rm{\Phi_{\rm{c}}}$, and the gas mass, $M_{\rm{g}}$ within
$3\,\rm{kpc}$.  The   black line illustrates  the  case  of  full  gas
consumption.  The red line stands for the case of an outflow. Finally,
the blue  line indicates the  evolution of a  self-regulated model. The
dotted black line delimits the star formation threshold density.}
\label{MLSfr} \end{figure}

Besides  this general  description    common to all  simulations,   we
identify three different major  regimes. We refer  to them  as ``full gas
consumption'', ``outflow'' and ``self-regulation''.
For each of them, Fig.~\ref{MLSfr} presents the evolution with time
of the star formation rate (SFR), the central gas density, the central
gravitational potential and the mass of  the gas within $3\,\rm{kpc}$ from the
galaxy center.


\subsubsection{Full gas consumption}\label{collapse}

In cases where  $c_\star$ is low  for a given $M_{\rm{i}}$, the energy
released by the supernova explosions is  unable to counterbalance the
radiative cooling.  As a consequence, the gas  keeps on sinking in the
galaxy   inner regions  and reaches   very high  densities.  Stars are
formed continuously   and     at  high  rate.   The    model    \#647
($c_\star=0.05$  and $M_{\rm{i}}=6.6\times   10^8\rm{M}_\odot$)     in
Figure~\ref{MLSfr}  provides  a clear  example  of this regime: steep
rise in   star formation rate and  central  gas density.  The chemical
enrichment of the resulting systems  is rapid, and their  metallicities
quickly exceed  the highest ones  measured in dSphs. These models were
not investigated further.


%
%


\subsubsection{Outflow}\label{outflow}


Stars can  be formed  at   slightly  lower densities  by    increasing
$c_\star$ at a  given initial mass, or by  decreasing the initial mass
at   fixed   $c_\star$.  This  is  sufficient   to   stop  the drastic
accumulation of  gas at the  center.  Nevertheless, the gas density is
still high, and star formation is very efficient.  When SNe explode, a
huge amount  of energy  is deposited  in the gas,   which  in turn  is
expelled from the galaxie's central regions.  In parallel, the central
potential increases (it is negative), primarily due to the ejection of
the the gas, but also due to the ensuing DM halo expansion.  The final
consequence   is a strong  outflow. 
A large fraction of the total gas mass is ejected beyond a radius of $3\,\rm{kpc}$, chosen
to be large enough compared to the stellar extent of the systems.
For clarity,  we illustrate this
regime with   M\#519 in Fig.\ref{MLSfr}, in  which  the outflow occurs
early in the galaxy    evolution: after $\sim 2\,\rm{Gyr}$, there   is
virtually no gas left.



\subsubsection{Self-regulation}\label{regulated}

Dwarf     galaxies are  formed     in  a  regime   of self-regulation,
characterized by   successive periods of  cooling  and feedback.  Such
intermittent  star formation    episodes  occurring  spontaneously   in
hydrodynamical simulations  have been  mentioned by \citet{stinson07}
and \citet{valcke08}.

\begin{figure*}
\resizebox{\hsize}{!}{\includegraphics[angle=0]{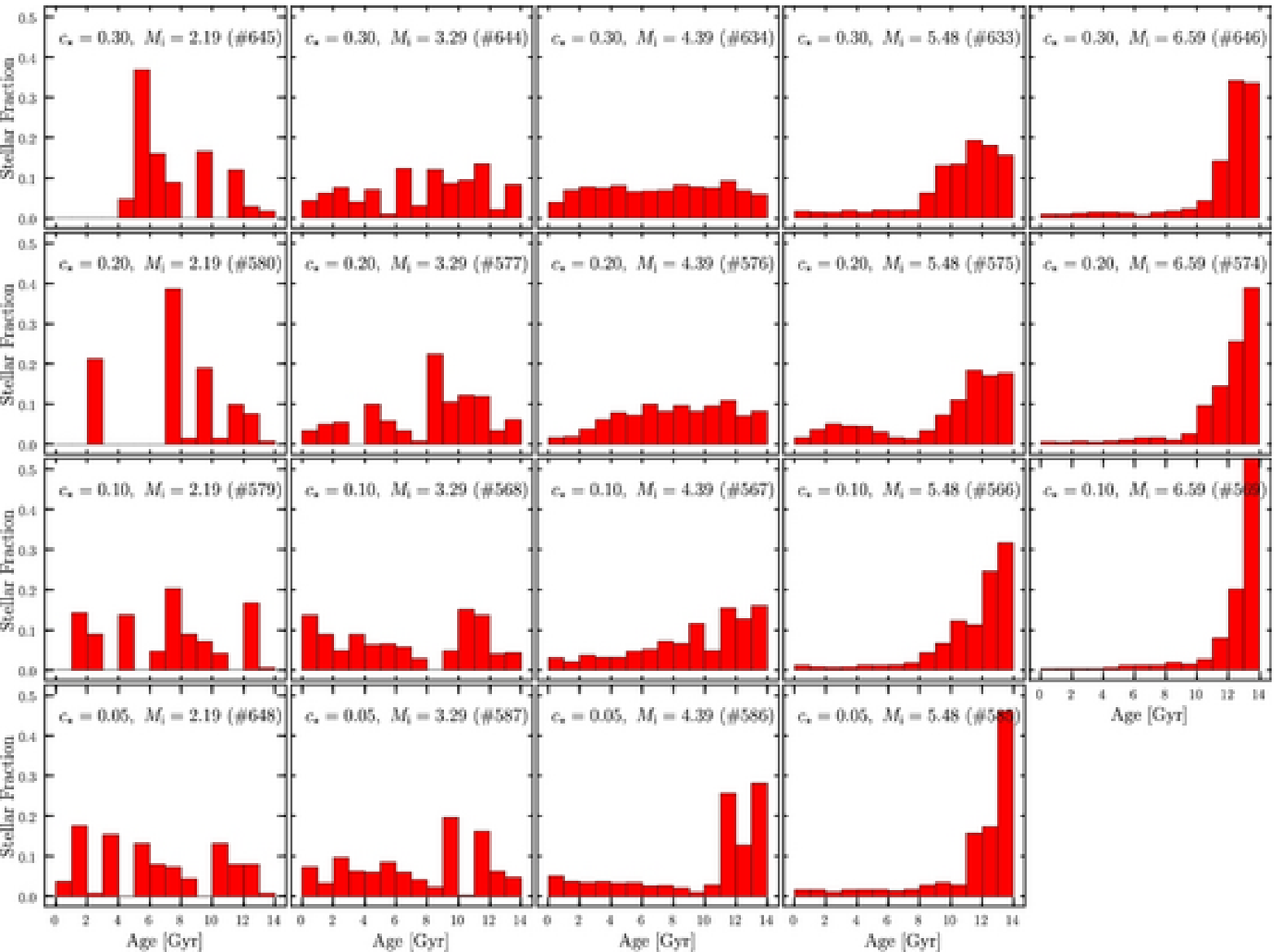}}
\caption{The distributions of stellar ages between $0$ and $14\,\rm{Gyr}$ for a
sub-sample of simulations with $f_{\rm{b}}=0.15$ and $r_{\rm{c}}=1$.}
\label{CstarvsMi2}
\end{figure*}

M\#576 in Fig.\ref{MLSfr} offers the example of an intermediate mass
self-regulated system ($M_{\rm{i}}=4.39\times 10^8\,\rm{M_\odot}$).
As usual, the first contraction of the gas leads to a peak in star
formation ($t=0.6\,\rm{Gyr}$).  The gas expelled by the supernova
feedback is diluted at densities below $\rho_{\rm{sfr}}$, and the star
formation stops.  As the gas particles cool, they become eligible to
star formation again, forming a new burst. Star formation occurs at
high frequency in M\#576 (periods between $100$ and $200\,\rm{Myr}$).
It produces a flat distribution of stellar ages, mimicking a nearly
continuous star formation rate (see Fig.~\ref{CstarvsMi2}).  We will
show later that the corresponding chemical signatures are also very
homogeneous.

Contrary to what has been observed by \citet{stinson07}, the
fluctuation of the SF is not strictly periodic.  However, we confirm
the influence of the total mass $M_{\rm{i}}$ on the duration of the
quiescent periods \citep{valcke08}.  Lower mass systems ($M_{\rm{i}}
\lessapprox 3\times 10^{8}\rm{M_\odot}$) are generally characterized
by star formation episodes separated by longer intervals, up to a few
Gyrs. These systems exhibit inhomogeneous stellar populations.



%
%


\subsection{Driving parameters}


%
\begin{figure*}
\resizebox{\hsize}{!}{\includegraphics[angle=0]{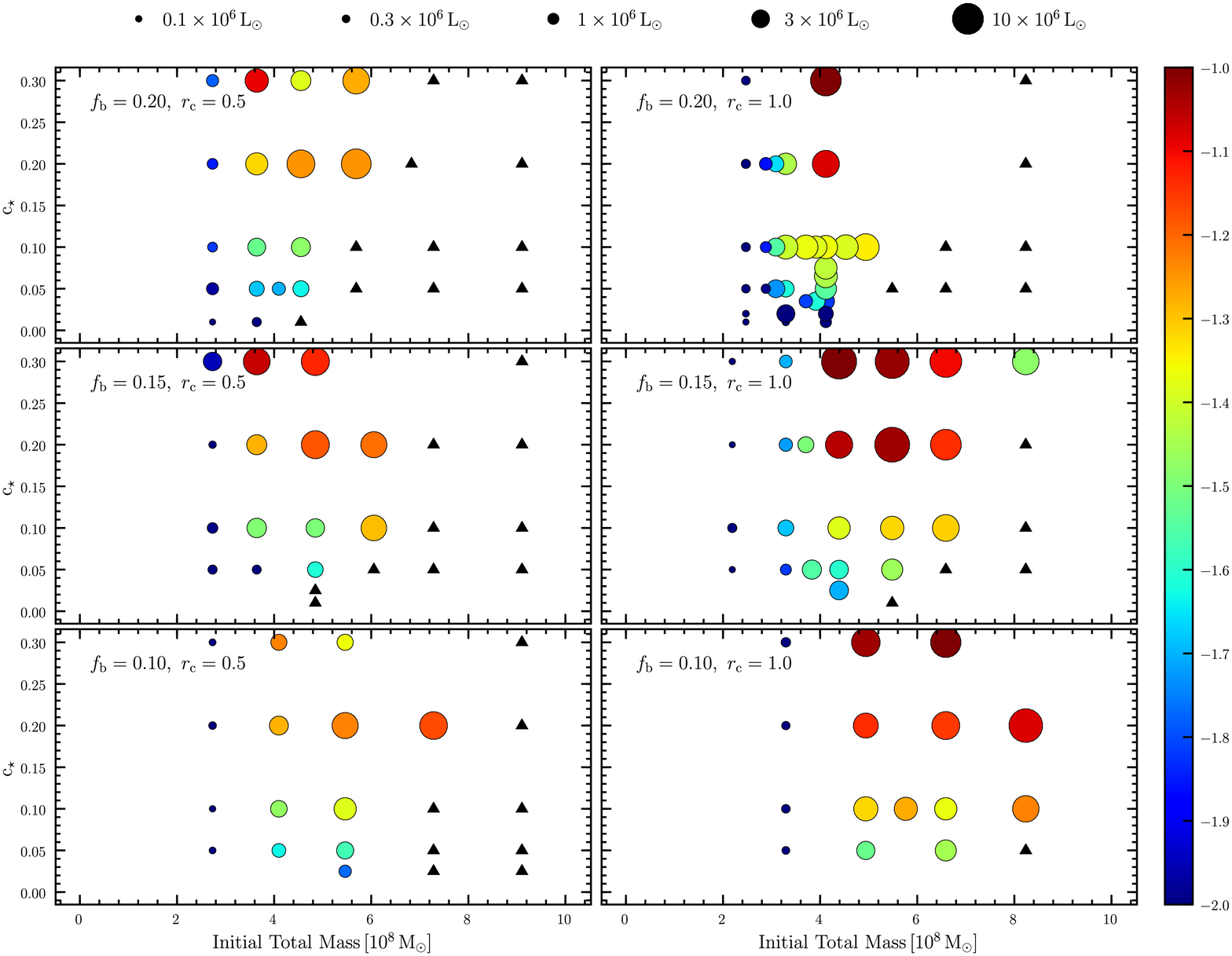}}
\caption{The effect of variation of the model {\rm initial} parameters. The full
  sample of simulations is shown. Each circle or triangle stands for
  one model.  The circle size is proportional to the final galaxy
  luminosity (see the top of the Figure), while colors code the final stellar metallicity (median
  of the metallicity distribution function), following the scale given
  at the right hand side of the diagrams.  Each panel encompasses
  simulations of identical baryonic fraction and initial core radius.
  The black triangles identify cases of full gas consumption.}
\label{CstarvsMi}
\end{figure*}

The description of the different  regimes of star formation histories
already points out   the importance of  both  the initial  total mass,
$M_{\rm{i}}$,  and the  star  formation parameter,  $c_\star$.  Figure
\ref{CstarvsMi2} presents  the stellar age  histograms for models with
$f_{\rm{b}}=0.15$ and $r_{\rm{c}}=1$.  From  bottom to top,  $c_\star$
increases by a factor  6.  From left  to right, $M_{\rm{i}}$ increases
by a factor 3.  The highest mass systems are characterized by a strong
predominance of the old stellar populations.  Decreasing the initial total
mass  extends the period of star formation, passing
progressively from a continuous to  a discrete distribution of stellar
ages.  The role of  $c_\star$ appears secondary, distributing slightly
differently the different peaks  of star formation  (position
and strength).

Fig.~\ref{CstarvsMi} summarizes the \numsim\, simulations in a diagram
of $c_\star$ and $M_{\rm{i}}$, for different core radii $r_{\rm{c}}$
and baryonic fractions $f_{\rm{b}}$.  Colors code the final galaxy
stellar metallicity \mfe, computed as the median of the distribution,
since it best traces the position of the metallicity peaks in the
observations.  The size of the circles is proportional to the final
stellar luminosity in the $V$-band, $L_{\rm{v}}$.
The small  black triangles  indicate  the simulations  that lead to
full gas consumption and have been stopped.  As described earlier, 
the latters result from a too small $c_\star$.
It can  be avoided, for our
purpose,   by   increasing  $c_\star$    or  decreasing  $M_{\rm{i}}$.
Self-regulated  systems with limited   outflow are  found for  smaller
$M_{\rm{i}}$.   These tendencies do not  dependent on $f_{\rm{b}}$ and
$r_{\rm{c}}$, which can only slightly  modify the interval of mass in
which a  particular regime is  valid.  For  a given $c_\star$,  \mfe\,
increases with $M_{\rm{i}}$  and similarly, for  a given $M_{\rm{i}}$,
\mfe\, increases  with $c_\star$.  At  very low mass, however, \mfe\, is
only  weakly influenced by $c_\star$. On  the contrary, the larger the
mass, the smaller $c_\star$ increase is needed to raise \mfe.

\begin{figure*}
\resizebox{\hsize}{!}{\includegraphics[angle=0]{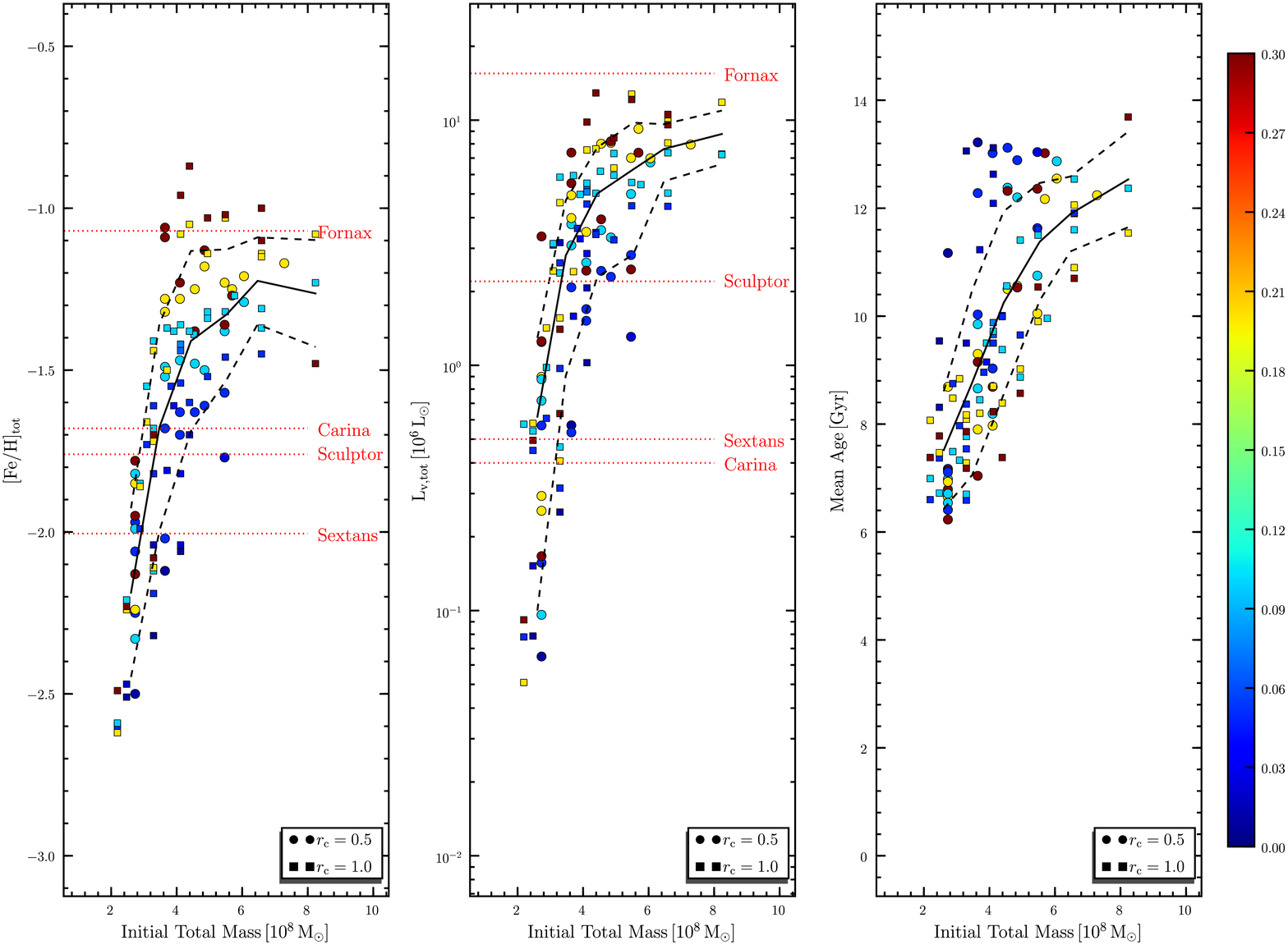}}
\caption{The metallicity, \mfe, luminosity in $V$-band,  $L_{\rm{v}}$,
and stellar mean age as  a function of  the galaxy's total initial mass,
after $14\,\rm{Gyr}$ of evolution.  The circles and squares correspond
to $r_{\rm{c}}=0.5$ and $r_{\rm{c}}=1$, respectively.  The colors encode
$c_\star$ following  the  scale at  the right-most panel of
the figure. The positions of our four targeted dSphs are indicated with
red dotted lines.
Plain line indicate the mean values, while dashed ones indicate the one sigma
deviation.}
\label{LFeAgevsMi} 
\end{figure*}

The left and middle panels of Fig.~\ref{LFeAgevsMi} display the final
galaxy stellar metallicity and stellar $V-$luminosity, respectively,
as a function of $M_{\rm{i}}$.  The Local Group dSphs luminosities
\citep{mateo98,grebel03} and mean metallicities (DART) are indicated
with horizontal dotted red lines.  The most outstanding result is that
changing $M_{\rm{i}}$ by a factor 4 translates to a change in
$L_{\rm{v}}$ by a factor 100. \mfe\, is varied by a factor $\sim$ 3.
By comparison, the influence of $c_\star$ on the galaxy properties
appears small. In any case, increasing $c_\star$ will also help
increasing both $L_{\rm{v}}$ and \mfe.  The consequence of varying
$M_{\rm{i}}$ is not linear.  At low initial mass, a small increase in
mass is sufficient to strongly increase $L_{\rm{v}}$ and \mfe, while at
larger initial mass, the relations saturate and a larger step in mass
is necessary. The mass-luminosity and metallicity-luminosity relations
will be discussed in the next section.

The right panel of Fig.~\ref{LFeAgevsMi} presents the relation between
the galaxy's mean stellar age and $M_{\rm{i}}$.  The influence of
$c_\star$ looks more linear than previously on $L_{\rm{v}}$ and \mfe.
As a matter of fact, we have seen in Fig.~\ref{CstarvsMi2} that it
plays a role in the stellar age distribution. At a given initial mass,
$c_\star$ determines the length of the star formation periods as well
as the interval between them.

As a conclusion, the above analyses stress the primordial impact
of the initial  total mass of the  systems.  Moreover, one can clearly
identify the  range of possible $M_{\rm{i}}$ leading to the formation
of  the Local Group  dSphs  as we observed  them  today. This range is
narrow,  e.g,   a  factor  2   centered  on  $M_{\rm{i}}  \ge  5\times
10^{8}\rm{M_\odot}$.


%
%


%





\section{Global relations}\label{global_relations}




dSph  galaxies   follow   luminosity-mass and   luminosity-metallicity
relations that are considered  as  cornerstones to understanding   their
formation and evolution
\citep{mateo98,wilkinson06,gilmore07,strigari08,geha08,kirby08}.


In the following discussion, we calculate all physical quantities
(luminosities, masses, abundances) within the radius $R_{\rm{L}}$
defined as the radius containing 90\% of a galaxy's total
luminosity. This choice is guided by the wish to reproduce as closely
as possible the observational conditions under which these quantities
are measured.  The classical dSphs (as opposed to newly discovered
faint ones) surrounding the Milky Way have tidal radii in the range
$\sim$ $0.5\,\rm{kpc}$ to $3\,\rm{kpc}$ \citep{irwin95}.  Fixing a
constant small aperture for all dSphs would underestimate both light
and mass of the largest systems.  Since dark matter does not
necessarily follow light, this would also bias the results.  The
observational estimates of the dSph total masses are based on stellar
velocity dispersions measured at galactocentric radii as large as
possible, thereby directly linked to the limits of the visible matter.
Although the farthest measurements do not always reach the galaxies'
tidal radii, their location is determined by severe drops in stellar
density, ensuring that the bulk of the galaxies' light is
enclosed. Consequently, we compare our models to the masses derived at
the outermost velocity dispersion profile point
\citep[e.g,][]{walker07, battaglia08, kleyna04}.  Ursa Minor is the
only exception to this rule. Its mass has been derived from its
central velocity dispersion \citep{mateo98}.

\begin{figure}
\resizebox{\hsize}{!}{\includegraphics[angle=0]{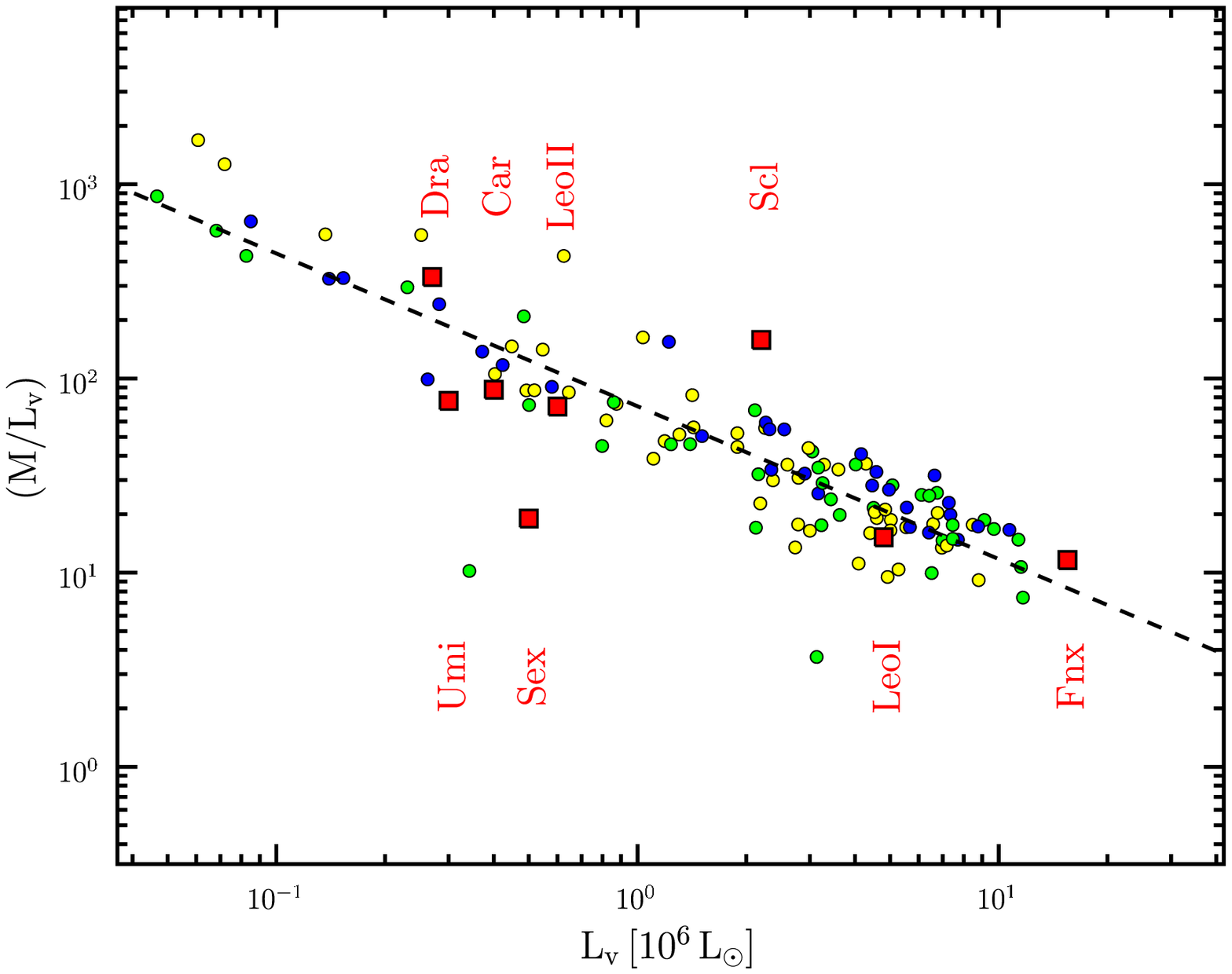}}
\caption{The galaxies' mass-to-light ratios, $M/L_{\rm{v}}$, versus
  their total luminosities  $L_{\rm{v}}$ in the $V$-band.  Each circle
  stands for  a  model.  Colors  encode   the three different  initial
  baryonic fractions   that  we   have  considered,   $f_{\rm{b}}=0.2$
  (yellow), $f_{\rm{b}}=0.15$   (green) and  $f_{\rm{b}}=0.1$  (blue).
  The red  squares represent the Milky Way  brightest dSphs.  The best
  fit to  our data  (Eq.~\ref{MLvsL_fit})  is indicated  by the dashed
  line.}
\label{MLvsL}
\end{figure}

\begin{figure}
\resizebox{\hsize}{!}{\includegraphics[angle=0]{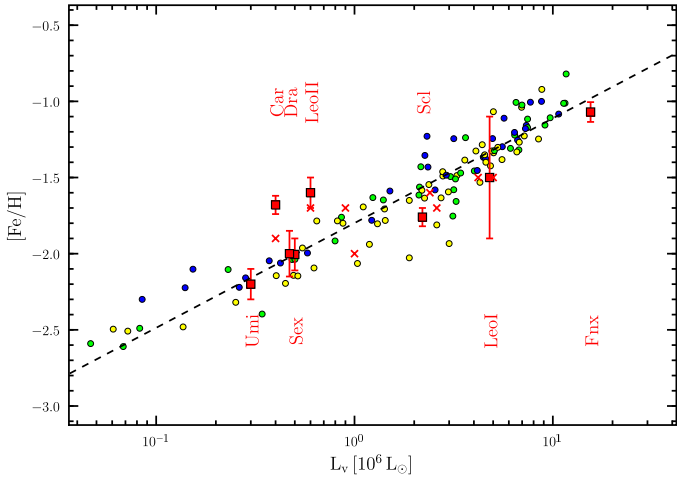}}
\caption{ The median of the galaxies' metallicity distributions, \mfe,
  as a function of their luminosities $L_{\rm{v}}$.  Each circle
  stands for a model.  Colors encode the three different baryonic
  fractions that we have considered, $f_{\rm{b}}=0.2$ (yellow),
  $f_{\rm{b}}=0.15$ (green) and $f_{\rm{b}}=0.1$ (blue).  The best fit
  to our data (Eq.~\ref{MLvsL_fit}) is indicated by the dashed line.
  The red squares represent the Milky Way's brightest dSph satellites,
  the red crosses represent the M31's ones. }
\label{FevsL}
\end{figure}

Figures ~\ref{MLvsL} and \ref{FevsL} display the relations of the
galaxies' mass-to-light ratios $M/L_{\rm{v}}$ and the median of the
stellar metallicity distributions \mfe, together with the total
luminosity of the model galaxies.
The observations are represented in red, with squares for the Milky
Way satellites and crosses for the others. In general, the values of
\mfe\ are taken from \citet{mateo98} when available or from
\citet{grebel03} otherwise. The mean metallicities of Carina, Fornax,
Sculptor, and Sextans are calculated from their metallicity
distributions \citep{helmi06,battaglia08,battaglia06}.The mean
metallicity of Leo II is derived from the metallicity distribution of
\citet{bosler07}. The luminosities are taken from \citet{grebel03},
with the exception of Draco \citep{martin08}.  The masses of Carina,
Fornax, Draco, Leo I and Leo II are computed by \citet{walker07}
inside $r_{\rm{max}}$.  The mass of Sextans corresponds to the upper
limit of \citet{kleyna04}, while the mass of Ursa Minor comes from
\citet{mateo98}.  Sculptor's $M/L_{\rm{v}}$ is taken from
\citet{battaglia08}.

Both $M/L_{\rm{v}}$  and \mfe\,  show very clear  log-linear
relations with $L_{\rm{v}}$:
	\begin{equation}
	\log_{10} \left( M/L \right) = -0.79 \log_{10} \left( L_{\rm{v}} / L_{\odot} \right) + 1.85,
	\label{MLvsL_fit}
	\end{equation}
and
	\begin{equation}
	\langle \rm{[Fe/H]}\rangle = 0.68 \log_{10} \left( L_{\rm{v}} / L_{\odot} \right) -1.8.
	\label{FeHvsL_fit}
	\end{equation}

Despite differences in  $c_\star$, $r_{\rm{c}}$ and $f_{\rm{b}}$,  all
simulations nicely reproduce the observations, with a reasonably small
scatter.  This stresses once  more that a dSph galaxy's total initial
mass drives most of its evolution.

%

%

%


\begin{figure}
\resizebox{\hsize}{!}{\includegraphics[angle=0]{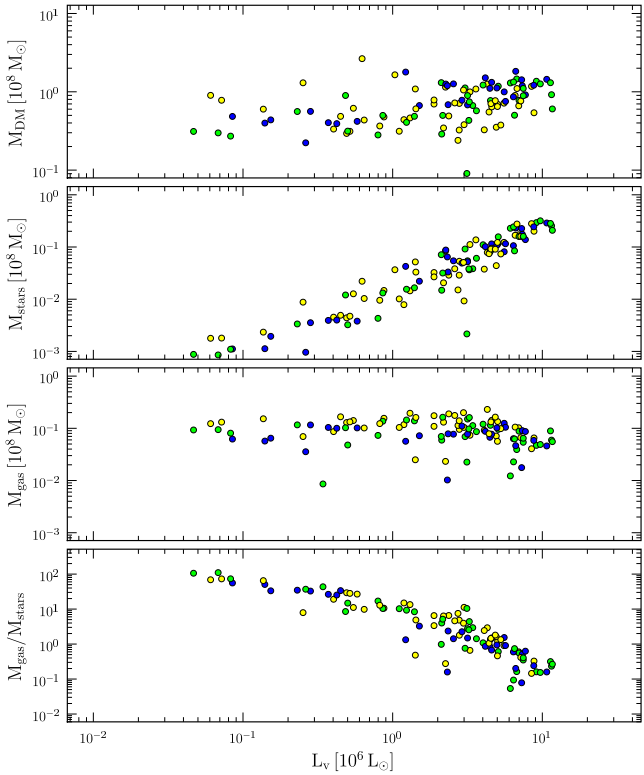}}
\caption{The relation  between  galaxy mass  and $V$-band luminosity  for the
the dark matter,  the  stars  and  the gas. 
The lower panel displays the gas over stars mass ratio. 
Colors encode  the three different baryonic  fractions that we have considered,
$f_{\rm{b}}=0.2$  (yellow),   $f_{\rm{b}}=0.15$ (green)  and  $f_{\rm{b}}=0.1$
(blue).}
\label{MvsL}
\end{figure}

To allow deeper insight into the building up of the
$M/L_{\rm{v}}-L_{\rm{v}}$ relation, Fig.~\ref{MvsL} distinguishes
between dark matter (DM), stars and gas. Colors encode the three
different initial baryonic fractions that we have considered,
$f_{\rm{b}}=0.2$ (yellow), $f_{\rm{b}}=0.15$ (green) and
$f_{\rm{b}}=0.1$ (blue).

Very naturally, stellar mass scales with the luminosity.  At fixed
luminosity, the dispersion in stellar mass is of the order of $5
\times 10^6$M$_{\odot}$.  In fact, a more appropriate way to look at
this panel is to consider the dispersion in luminosity at fixed
stellar mass, since the dispersion in luminosity is a direct
consequence of various distributions in stellar ages and
metallicities.  This dispersion, of the order of $
10^6\,\rm{L_\odot}$. It increases slightly for larger masses, for
which star formation can last longer, inducing a larger number of
possible age/metallicity combinations.

As already discussed in Fig.~\ref{LFeAgevsMi}, whilst $L_{\rm{v}}$
spans nearly 3 orders of magnitude, the mass of dark matter varies
little.  This variation is much less than one order of magnitude
inside $R_{\rm{L}}$ and is mostly due to the dispersion among the
models.  A common mass scale (inside $R_{\rm{L}}$), around $1-5 \times
10^7\,\rm{M_\odot}$, for the dSph total masses seems also favored by
the observations, although an exact value for this limit is difficult
to ascertain, given the large uncertainties of the mass estimates in
general \citep{mateo98, gilmore07, strigari08}.

Interestingly, one can now witness the effect of varying the initial
baryonic fraction. Galaxies with $L_{\rm{v}} < 3\times
10^6\,\rm{L_\odot}$ exhibit identical DM halo masses, whatever
$f_{\rm{b}}$, while for larger luminosities, the models of lowest
$f_{\rm{b}}$ require larger halo masses in order to generate a similar
quantity of stars.  For $f_{\rm{b}}=0.20$, the DM halo mass is
constant over the whole luminosity range.  This demonstrates that
while dark matter plays a crucial role in confining the gas, the
amount of the latter is also important for the most massive dSphs: it
must reach a critical amount to enable their formation.

The third panel of Fig.~\ref{MvsL} displays the mass of residual gas
after $14\,\rm{Gyr}$.  It constitutes a very small fraction of the
total mass, and is therefore undetectable at the level of the scaling
relations.  Quite interestingly, its amount is of the order of the HI
mass observed in dwarf irregular galaxies (dIs) \citep{grebel03} and
is weakly dependent on the total luminosity.  Nevertheless, there is a
non-intuitive tendency for the most luminous galaxies ($>5\times
10^6\,\rm{L_\odot}$) to exhibit less gas than the rest of the systems
on the sequence.  However, the most massive galaxies succeed in
retaining most of their gas despite the supernova explosions. Less
than 60\% of their gas is ejected, while this fraction lies between
70\% and 90\% for less massive systems, in agreement with the results
of \citet{valcke08}.  However, the star formation efficiency is also
higher in more massive galaxies.  As a result, the gas consumption
counterbalances the presence of the large gas reservoir. We will come
back to this in Section~\ref{discussion}.

Since the gas mass is very much constant at around
$10^7\,\rm{M_\odot}$, the smallest galaxies have proportionally more
gas than the massive ones (see bottom panel of Fig.~\ref{MvsL}).  
Below $ 3 \times 10^6\,\rm{L_\odot}$,
galaxies have more gas left than they have formed stars. Their gas to
stellar mass ratios reach 100 at the faint end of the dSph model
sequence.  Meanwhile, stars dominate over gas by a factor 5 for the
most luminous galaxies.
  
As a conclusion, the lessons to be taken from Fig.~\ref{MvsL} are
twofold: i) all dSphs are clearly dominated by dark matter.  The lower
the mass of the system, the lower the final baryonic fraction.  ii)
all model galaxies present an excess of gas at the end of their
evolution, as has been found in similar studies
\citep{marcolini06,marcolini08,stinson07,valcke08}.  As demonstrated
earlier, dSphs cannot originate from smaller amounts of gas (for a
given final luminosity, metallicity and age).  It is necessary to
initiate the star formation in the observed proportions.  However, it
is not yet clear how much of this gas must be kept in the subsequent
phases of the galaxy evolution.  It is clear that the excess of gas,
observed in models in isolation, must in reality have been stripped
some time during the galaxy evolution.

The quantity of gas falls in the range of HI mass observed in dIs,
always found further away than dSphs from their parent galaxies,
thereby bringing another piece of evidence for the role that
interactions might play in its removal. It might also be achieved in a
hierarchical formation framework, for which the gravitational
potentials are initially weaker.


As we have just shown, the global scaling relations are reproduced by
our model with impressive ease.  In turn, this conveys the idea that
the global scaling relations do not form a very precise set of
constraints.  They do not reflect the diversity in star formation
histories that we have illustrated.  In order to understand the
individual histories of the Local Group dSphs, one definitely needs to
go one step further and consider their chemical abundance patterns, as
well as the information that color-magnitude diagrams provide on the
stellar age distributions.


\section{Generic models}\label{generic_models}



We will now select and discuss a series of generic models
reproducing the properties of four Milky Way dwarf spheroidals,
Carina, Fornax, Sculptor and Sextans.  The choice of these models is
based on four observational constraints: 
\begin{itemize}
\item[1)] The dSph total luminosity, which scales with the total
  amount of matter involved in the galaxy star formation history.
\item[2)] The metallicity distribution which traces the star formation
  efficiency.
\item[3)] The chemical abundance patterns, in particular that of the
  $\alpha$-elements, that determines the length and efficiency of the
  star formation period together with its homogeneity.  We use
  magnesium for the $\alpha$-elements. [Fe/H] and [Mg/H] are
  derived from high resolution spectroscopy in the central regions of
  the galaxies \citep[Hill et al in preparation; Letarte et al. in
    preparation;][]{venn05,letarte07,shetrone03,koch08}.
\item[4)] The stellar age distributions. They complement the above
  constraints with information on possible series of bursts
  \citep{smecker-hane96,hurley-keller98,coleman08}.

\end{itemize}

Table~\ref{table2} lists the initial parameters of our generic models.
Table~\ref{table3} lists their final properties, namely the stellar
luminosity in the $V$-band, $L_{\rm{v}}$, the mass-to-light ratio,
$M/L_{\rm{V}}$, the gas mass, $M_{\rm{gas}}$, the median metallicity,
\mfe, the fraction of stars with [Fe/H] lower than $-3$
and, finally, the stellar age distribution divided in 3 bins: younger
than $4\,\rm{Gyr}$, between $4$ and $8\,\rm{Gyr}$ and older than
$8\,\rm{Gyr}$.

\begin{figure}
\resizebox{\hsize}{!}{\includegraphics[angle=0]{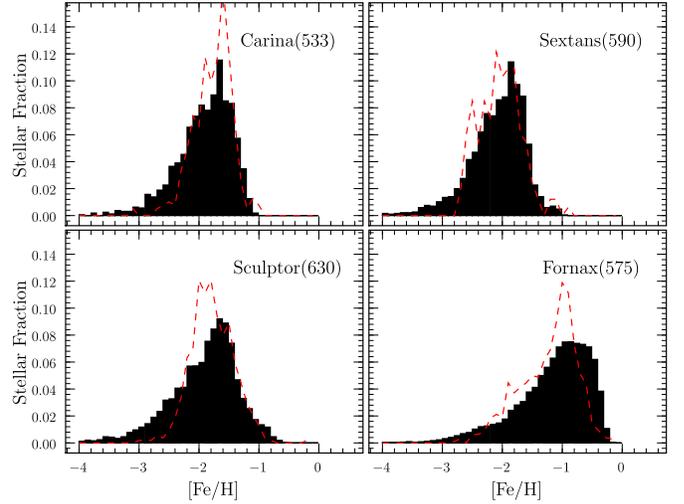}}
\caption{The stellar [Fe/H] distributions of the four generic models
  reproducing Carina, Sextans, Sculptor and Fornax.  The model outputs
  are shown in black and the observations with red dashed lines. The
  stellar fraction is defined as the number of stars in each bin
  divided by the total number of stars.}  \label{NvsFe} \end{figure}

\begin{figure}
\resizebox{\hsize}{!}{\includegraphics[angle=0]{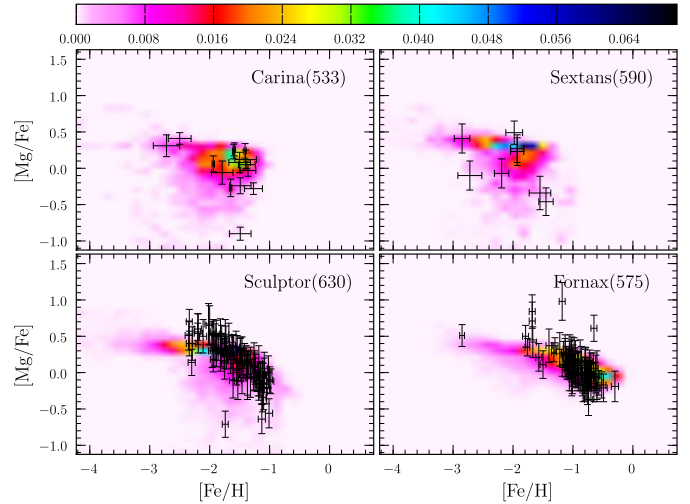}}
 \caption{ The stellar [Mg/Fe] vs [Fe/H] diagrams for the four generic
   models reproducing Carina, Sextans, Sculptor and Fornax.  The
   observations are sketched with black dots.  Their error bars are
   shown. Colors encode the fraction of stars as indicated by the bar
   at the top.} \label{MgFevsFe} \end{figure}

\begin{figure}
\resizebox{\hsize}{!}{\includegraphics[angle=0]{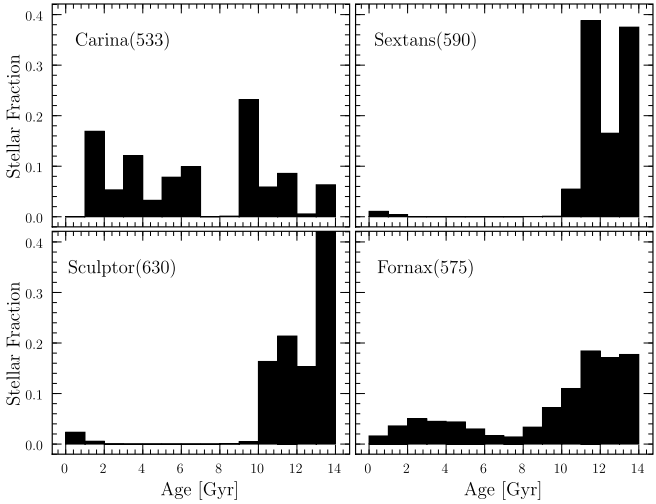}}
\caption{The stellar age distributions of the four generic models.
  The stellar fraction is defined as the number of stars in each bin
  divided by the total number of stars.}
  \label{NvsAge}
  \end{figure}

\begin{figure}
\resizebox{\hsize}{!}{\includegraphics[angle=0]{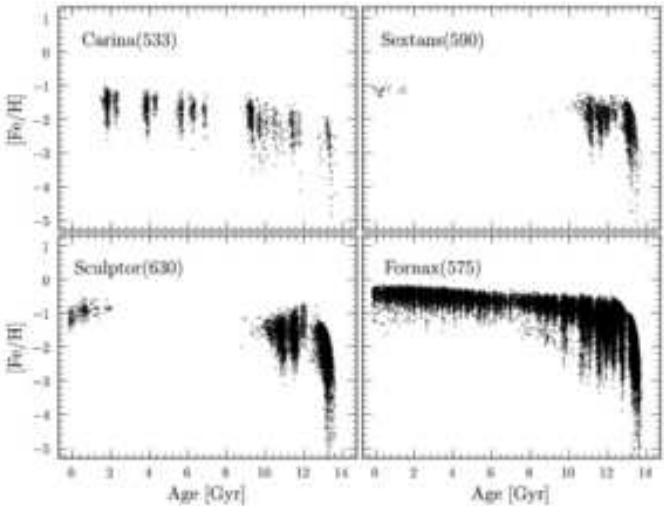}}
\caption{The age-metallicity relations of the four generic models. Each dot 
corresponds to a stellar particle.}
  \label{FevsAge}
  \end{figure}

Fig.~\ref{NvsFe}, \ref{MgFevsFe} and \ref{NvsAge} display the stellar
metallicity distribution, the [Mg/Fe] vs [Fe/H] diagrams and the
stellar age histograms, respectively, after $14\,\rm{Gyr}$ of
evolution. Fig.~\ref{FevsAge} shows the stellar age-metallicity relations.

\begin{figure}
\resizebox{\hsize}{!}{\includegraphics[angle=0]{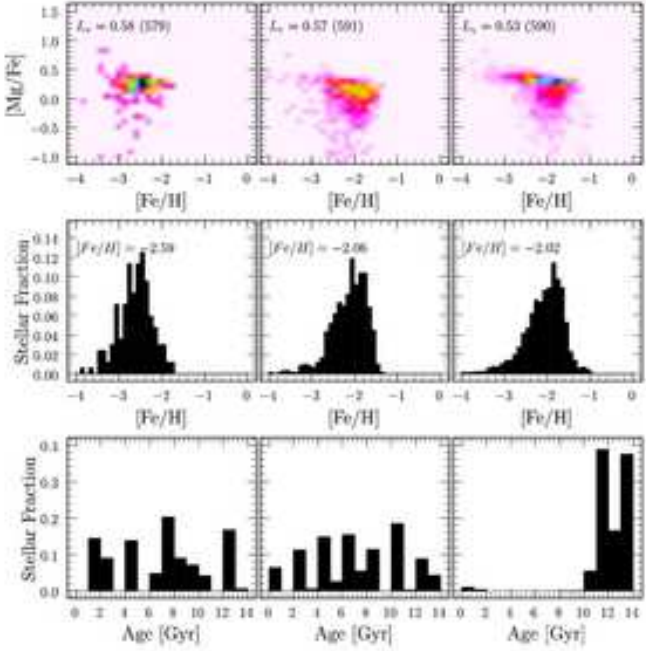}}
\caption{The variation in [Mg/Fe] vs [Fe/H], [Fe/H] distribution 
and stellar   age distribution for 3  models   of similar luminosities
($L_{\rm{v}}\cong 0.5\times 10^6\,\rm{M_\odot}$) and identical initial
baryonic   fraction ($f_{\rm{b}}=0.15$).    The   stellar fraction  is
defined as the number of stars in each bin divided by the total number
of stars.}  
\label{L0.4b} \end{figure}

\begin{figure}
\resizebox{\hsize}{!}{\includegraphics[angle=0]{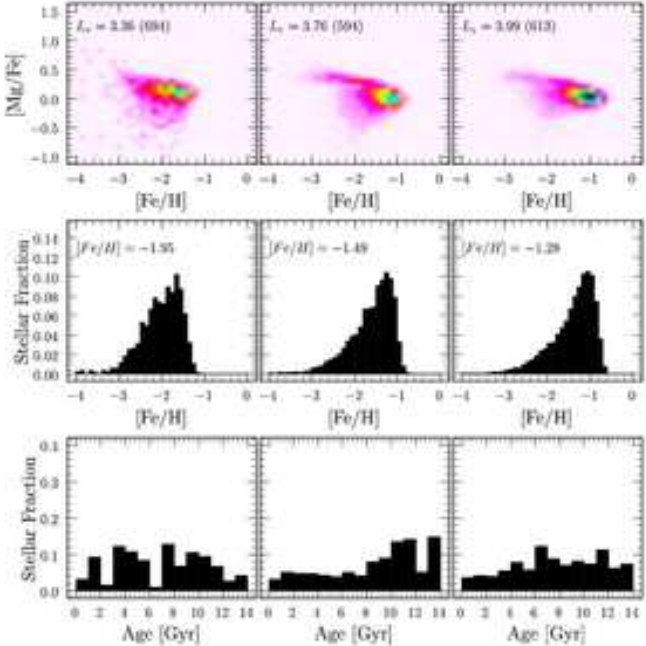}}
\caption{Same as  Fig.~\ref{L0.4b}  but for a brighter set of model
galaxies      ($L_{\rm{v}}\cong 3.7\times       10^6\,\rm{M_\odot}$)}
\label{L2.3b}
\end{figure}

We do not try to match exactly all properties of our target dSphs.
Instead, we select the four most satisfying models in our sample of
\numsim.  More specifically, we allow a freedom of a factor 2 in
luminosity and a shift of a few tenths of dex in peak [Fe/H].  It is
beyond doubt that we could fine-tune $M_{\rm{i}}$, $c_\star$,
$r_{\rm{c}}$ and $f_{\rm{b}}$ to exactly reproduce the
observations. However, this is beyond our scope, which remains to test
the hypothesis of a formation framework common to all dSphs.  The
dependence of the results on the model parameters is illustrated in
Fig.~\ref{L0.4b} and \ref{L2.3b}.  We selected 6 models at low
($L_{\rm{v}}\cong 0.5\times 10^6\,\rm{M_\odot}$) and high
($L_{\rm{v}}\cong 3.7\times 10^6\,\rm{M_\odot}$) luminosities.  The
baryonic fraction is fixed each time, and only $c_\star$ and
$M_{\rm{i}}$ vary. One sees that around a given fixed core of observed
characteristics (e.g, luminosity, metallicity), we could run the
models on a finer grid of parameters in order to match the galaxy
properties in all details. This means, for example, to exactly reflect
the stellar age distribution, or the spread in abundance ratios.


\subsection{Global evolution}

\begin{figure*}
\resizebox{\hsize}{!}{\includegraphics[angle=0]{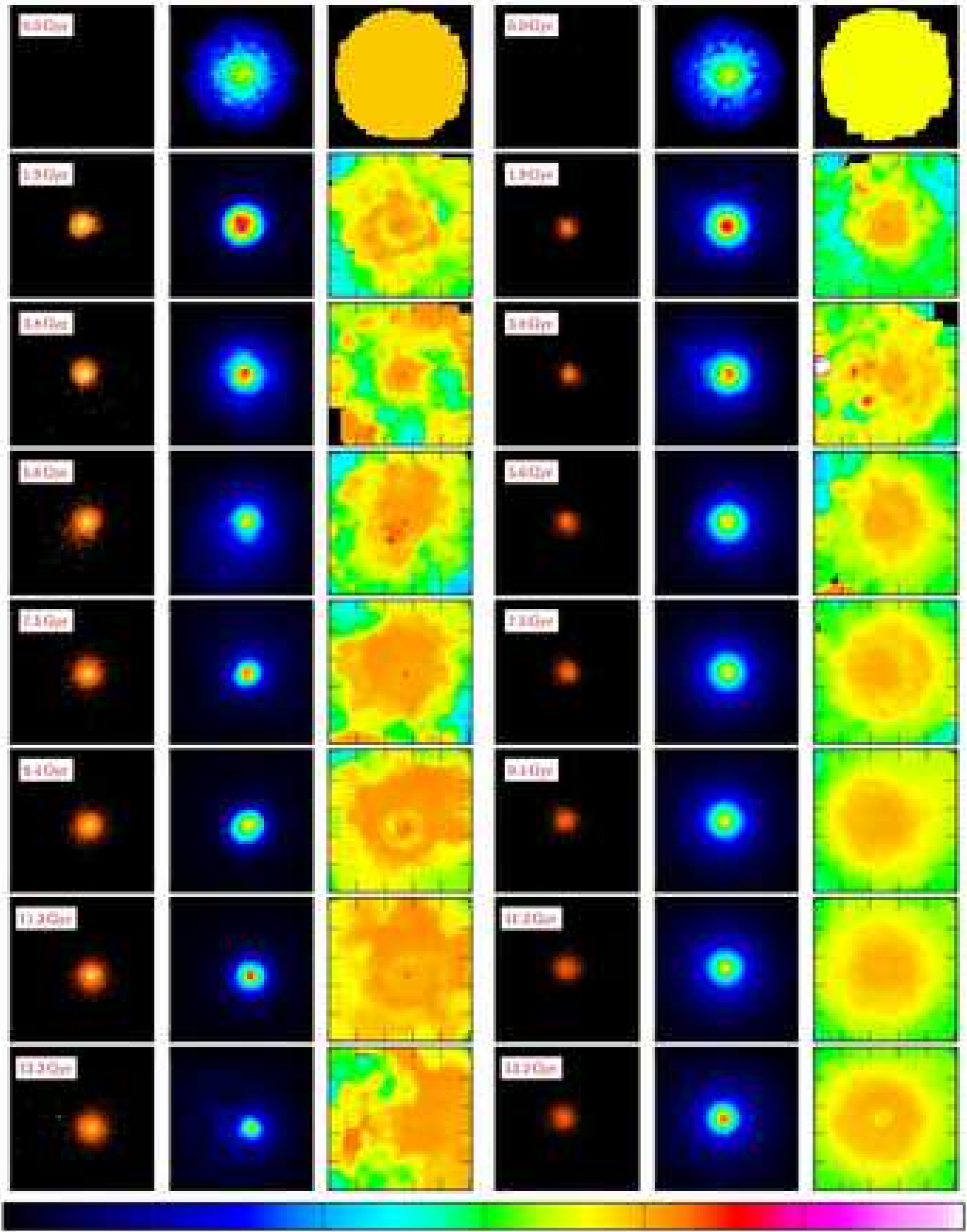}}
\caption{{\rm 
Stellar, gas and temperature evolution of the models \#575 (Fornax) and
\#630  (Sculptor) as a function of time.  
Each model is illustrated with three columns: 
The first column presents the stellar surface brightness in log scale, between $0$ and $0.4\,\rm{L_\odot}/kpc^2$,
the second one, the gas surface density in log scale, between $0$ and $1.5\times 10^7\,\rm{M_\odot}/kpc^2$,
the third one gives the log of the gas temperature, between $10^2$ and $10^5\,\rm{K}$. 
While intervals are different for the gas surface density and temperature, the principle of the
color coding remains similar: white-red colors for high values and blue-black colors for low values,
as indicated by the horizontal bar at the bottom  of the figure.
The  size of   each panel is $20\times   20\,\rm{kpc}$.
}}

\label{evolution12}
\end{figure*}

Fig.~\ref{evolution12} and \ref{evolution34} show the stellar surface
density, the gas surface density and the gas temperature between $0$
and $14\,\rm{Gyr}$ for our four selected generic models. Not only do
these reproduce the properties of Carina, Fornax, Sculptor and Sextans
individually, but they also depict the full spectrum of evolutions
seen in our models.  The size of each panel is $20\times
20\,\rm{kpc}$.

All gas particles initially share the same temperatures, corresponding
to the galaxy virial temperature, as given by Eq.~\ref{T_gas}.  As
soon as the simulations start, the gas looses energy by radiative
cooling.  Consequently, the gas density increases in the galaxy
central regions.  Red areas in the gas density maps identify densities
larger than $\rho_{\rm{sfr}}$, i.e., they mark regions where gas
particles may be eligible for star formation.  When this is indeed the
case, the newly formed stars are traced by their high brightness in
the stellar density maps.

The SN explosions induce temperature inhomogeneities.  Indeed, the
heated central gas expands and generates a wave propagating outwards.
If the SN feedback dominates the cooling, the gas is diluted and the
red color vanishes from the center of the gas density maps: star
formation is quenched.  Such quiescent periods are characterized by a
nearly constant and homogeneous central temperature $\sim
10^4\,\rm{K}$, the temperature at which the radiative cooling
counterbalances the adiabatic heating.  When gas has sufficiently
cooled, it can condense again, and star formation is induced again.
These periods of star formation and quiescence alternate at low or
high frequency, depending on the mass of the galaxy. For low-mass
systems, the cooling time is of the order of several Gyrs, it is much
shorter for more massive ones. We will now see how each of these cases
translate into dSph stellar population properties.


\subsection{Sextans}

\begin{figure*}
\resizebox{\hsize}{!}{\includegraphics[angle=0]{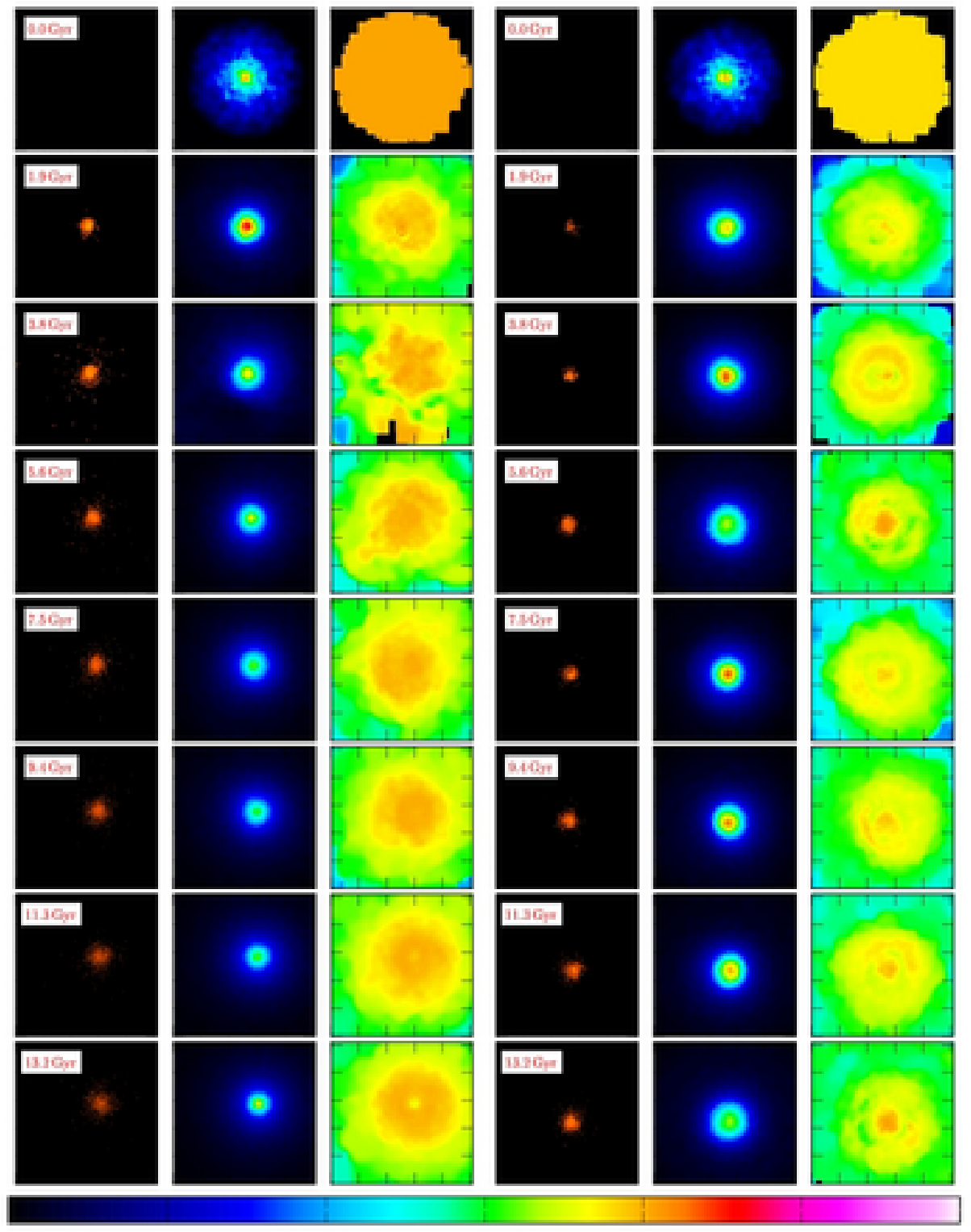}}
\caption{Same as Fig.~\ref{evolution12} but for the models \#590 (Sextans)  and \#533 (Carina).}
\label{evolution34}
\end{figure*}

The stellar population of Sextans is dominated by stars older than
$10\,\rm{Gyr}$ \citep{lee03}.  With its mean metallicity $\langle
\rm{[Fe/H]}\rangle = -2.05$ and a $V$-band total luminosity of $0.5
\times 10^6\,\rm{L_\odot}$, it falls exactly on our
luminosity-metallicity relation (Fig.~\ref{FevsL}), and its properties
are reproduced by the model \#590 ($L= 0.53\times 10^6\,\rm{L_\odot}$,
$\langle \rm{[Fe/H]}\rangle=-2.02$ ) that experiences an outflow.
Sextans' evolution is dominated by an early period of star formation
lasting about $ 3\,\rm{Gyr}$. The gas surface density map of
Fig.~\ref{evolution34} shows the dense central region at the origin of
the star formation burst ($t=1.9\,\rm{Gyr}$).  At that time, a small
but bright stellar system is already formed.  After this period, the
gas is expelled and diluted.  No star can form until the last
$\rm{Gyr}$, when the gas has sufficiently cooled down to fulfill the
star formation criteria.  This last episode of star formation is an
artifact of the gas retained by our model, as discussed in
Section~\ref{global_relations}.  It is however negligible compared to
the bulk of the population and does not influence Sextans' properties.

Fig.~\ref{MgFevsFe} shows that the bulk of Sextans model stars are
located at [Mg/Fe] $\cong 0.3$, with however a noticeable dispersion
at lower values.  The dispersion originates from the uneven
intensities of the star formation peaks.  They create regions with
diverse levels of chemical enrichment. When the intensity of star
formation rises again after a period of semi-quiescence, the ejecta of
new SNe II are mixed with material of low $\alpha$-element abundance.
Refinement of the model would require a larger sample of observed
stars at high resolution, particularly to estimate the statistical
significance of the dispersion in [Mg/Fe].

\subsection{Carina}

Carina seems to have experienced a complex evolution characterized by
episodic bursts of star formation, with at least three major episodes
at around $15$, $7$ and $3\,\rm{Gyr}$ , and a period of
quiescence between $7$ and $3\,\rm{Gyr}$
\citep{smecker-hane96,hurley-keller98}.   
Our set of  simulations reveal that episodic bursts of star formation are
intrinsic features of the  self-regulated  low initial   masses
systems   ($M_{\rm{i}}  \lessapprox  3\times  10^8\,\rm{M_\odot}$, see
Fig.~\ref{CstarvsMi2}).    Shortly   after  a star   formation episode,
supernovae explode, gas is heated and expands.  Because its density is
low, its   cooling time is of the   order of several Gyrs.   
This series  of well spaced-out  SF peaks translate into dispersion in
[Mg/Fe] at  fixed [Fe/H].   This  spread is more  accentuated than for
Sextans, due to the  extended star formation  and the longer intervals
between bursts.

Fig.~\ref{FevsL} shows that Carina, with $\langle
\rm{[Fe/H]}\rangle=-1.82$ and $L_{\rm{v}} = 0.72\times
10^6\,\rm{M_\odot}$, falls above the relation [Fe/H]-$L_{\rm{v}}$
defined by our models.  As already mentioned, we give priority to
constraints provided by the spectroscopic data, and allow some
flexibility in $L_{\rm{v}}$. Under these conditions, model \#533
($M_{\rm{i}} = 2.7\times 10^8\,\rm{M_\odot}$) provides a very
reasonable fit to Carina's properties.  The three major SF episodes
are reproduced: 45\% of the stars have ages between $9$ and
$15\,\rm{Gyr}$, 29\% have $4$ to $8\,\rm{Gyr}$ and 22\% are younger
than $3\,\rm{Gyr}$.  Whilst Carina is somewhat less luminous than
Sextans, its mean metallicity is higher.  This is due to a lower
initial mass, locking less matter in stars, but a longer period of
star formation, obtained by a slightly higher star formation
parameter, which avoids outflows.

\subsection{Sculptor}

The Sculptor dSph has formed stars early, over a period of a few Gyrs.
No significant intermediate age population has been found, hence
excluding star formation within the last $ \sim 5\,\rm{Gyr}$.  The
evidence for low $\alpha$-element enhancement agrees with an extended
star formation and self-enrichment over a period of at least
$2\,\rm{Gyr}$ \citep[e.g.][]{babusiaux05, shetrone03, tolstoy03}.

Sculptor's properties are well reproduced by model \#630 ($L=
2.86\times 10^6\,\rm{L_\odot}$, $\langle \rm{[Fe/H]}\rangle=-1.82$).
The outstanding feature of Sculptor, as compared to the two previous
dSph, is the small dispersion in the [Mg/Fe] vs [Fe/H] diagram
(Fig.~\ref{MgFevsFe}).  This is guaranteed by a strong initial star
formation, followed by subsequent episodes of lower, but smoothly
decreasing intensities, ensuring the chemical homogeneity of the
interstellar medium.

\subsection{Fornax}

Fornax is obviously the most challenging case of dSphs.  It is the
most luminous of all ($15.5\times 10^6\,\rm{L_\odot}$), it is
metal-rich ($\langle \rm{[Fe/H]}\rangle=-1.07$) and has experienced
multiple periods of star formation.  Indeed, in addition to very old
stars, Fornax reveals a dominant intermediate-age population, as well
as stars of $\sim 3-4\,\rm{Gyr}$ \citep{coleman08}.

Fornax is nicely reproduced by the self-regulated model \#575.  Stars
are formed in a high frequency series of short bursts, lasting between
$200$ and $500\,\rm{Myr}$.  The amplitude of these bursts
progressively decreases with time, until $\sim 7\,\rm{Gyr}$.
Thereafter, part of the gas that had been expelled by previous SN
explosions, has sufficiently cooled to restore star formation, at $
\sim 3\,\rm{Gyr}$, similar to what is observed. The succession of the
short bursts of similar intensities mimics a continuous star formation
and leads to an efficient and homogeneous chemical enrichment
($\langle \rm{[Fe/H]}\rangle=-1.03$).

We reach a luminosity $L= 12.6 \times 10^6\,\rm{L_\odot}$, slightly
below the observation of Fornax.  However, just as in the case of
Carina, the exact value of the $V$-band luminosity is very sensitive
to the exact amount of stars formed in the last Gyrs and may easily
vary.  Stars form up to the end of the simulation.  This is consistent
with the observational evidence of a small number of $100\,\rm{Myr}$
stars \citep{coleman08}.

Figure~\ref{FevsAge} show the stellar age-metallicity relations of our
four  generic models.   In all  cases, we are   far from a one  to one
correspondence between age and metallicity, although below [Fe/H]=-2.5
(but more safely below [Fe/H]=$-3$), stars are generally older than 10
Gyr. The dispersion  in metallicity is in  the  range 0.5 to 1  dex at
ages younger than 10 Gyr.

\begin{figure}
\resizebox{\hsize}{!}{\includegraphics[angle=0]{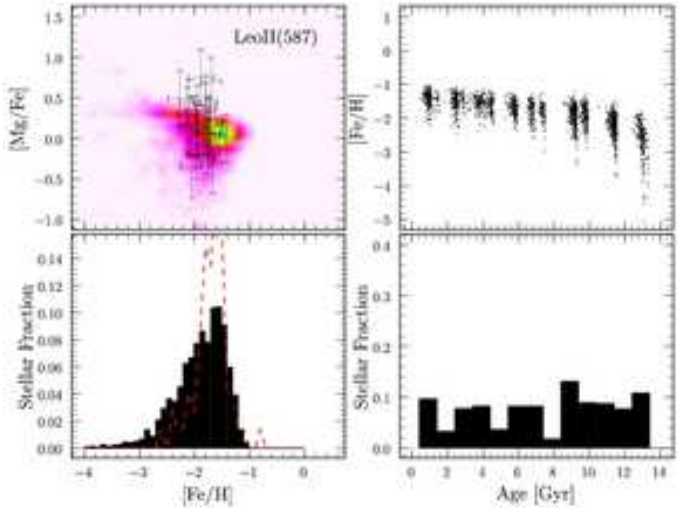}}
\caption{The  [Mg/Fe] vs [Fe/H] diagram, the stellar [Fe/H]
distribution, the age-metallicity  relation and the global stellar age
distribution of the model \#587,  compared to the observations of Leo
II}
\label{leoII}
\end{figure}

\begin{table}
    \begin{tabular}{llccccc}
    \hline
    \hline
  dSph &  \# &  $r_{\rm{c}}$  & $f_{\rm{b}}$ & $c_\star$ & $\rho_{\rm{c,tot}}$  & $M_{\rm{tot}}$ \\
   &   &  $[\rm{kpc}]$       &              &           & $[10^8\,\rm{M_\odot/kpc^3}]$   & $[10^8\,\rm{M_\odot}]$ \\

    \hline
    \hline
Carina   &   $533$ & 0.5 & 0.20 & 0.100 & 1.20e-03 & 2.73     \\
Leo II   &   $587$ & 1.0 & 0.15 & 0.050 & 4.00e-04 & 3.29     \\
Sextans  &   $590$ & 0.5 & 0.15 & 0.050 & 1.60e-03 & 3.64     \\
Sculptor &   $630$ & 1.0 & 0.20 & 0.035 & 5.00e-04 & 4.12     \\
Fornax   &   $575$ & 1.0 & 0.15 & 0.200 & 6.66e-04 & 5.48     \\

    \hline
    \end{tabular}
    \caption[]{Initial parameters of the five  generic models.}
    \label{table2}
\end{table}

\begin{table*}
    \begin{tabular}{llcccccccc}
    \hline
    \hline
  dSph &  \# &  $L_{\rm{v}}$ & $M/L_{\rm{V}}$ & $M_{\rm{gas}}$ & $\langle \rm{[Fe/H]}\rangle$& $\rm{[Fe/H]}<-3.0$ & $\le 4\,\rm{Gyr}$ & $[4,8]\,\rm{Gyr}$ & $\ge 8\,\rm{Gyr}$\\
   &   &   $[10^6\,\rm{L_\odot}]$ & & $[10^8\,\rm{M_\odot}]$ & dex & \% & \% & \% & \%\\
    \hline
    \hline
Carina &    $533$  &  0.72 & 231 &  0.28 & -1.82 &   2.9 &  26.1 &  29.3 &  44.6\\ 
Leo II &    $587$  &  0.97 & 219 &  0.33 & -1.82 &   2.9 &  24.4 &  26.4 &  49.2\\
Sextans &   $590$  &  0.53 & 390 &  0.17 & -2.02 &   5.2 &   1.7 &   0.0 &  98.3\\ 
Sculptor &  $630$  &  2.86 &  83 &  0.32 & -1.82 &   5.6 &   3.6 &   0.0 &  96.4\\ 
Fornax &    $575$  & 12.76 &  27 &  0.12 & -1.03 &   1.2 &  14.4 &  10.6 &  75.0\\ 

    \hline
    \end{tabular}
    \caption[]{The final properties of the five selected  generic models. 
    ($L_{\rm{v}}$)				: stellar luminosity in $V$-band,
    ($M/L_{\rm{V}}$)				: mass-to-light ratio,
    ($M_{\rm{gas}}$) 				: gas mass,
    ($\langle \rm{[Fe/H]}\rangle$)		: median of the metallicity distribution function,
    ($\rm{[Fe/H]}<-3.0$)			: fraction of stars having a metallicity lower than $-3$,
    ($\le 4\,\rm{Gyr}$)				: fraction of stars  younger than $4\,\rm{Gyr}$,
    ($[4,8]\,\rm{Gyr}$)				: fraction of stars having ages between $4$ and $8\,\rm{Gyr}$,
    ($\ge 8\,\rm{Gyr}$)				: fraction of stars older than  $8\,\rm{Gyr}$. 
    }
    \label{table3}
\end{table*}

\subsection{Leo II}

While writing up our results, the abundance ratios in a sample of 24
stars in Leo II have been published \citep{shetrone09}.  We keep Leo
II separated from the other individual case analyses, since its
abundance ratios have been obtained from lower resolution spectra and
therefore have larger uncertainties.

Fig.~\ref{leoII} illustrates how model \#587 can reproduce the star
formation history of Leo II.  The observed metallicity distribution,
derived from $\sim 100$ stars, is taken from \citet{bosler07}, and
[Mg/Fe] from \citep{shetrone09}.  Leo II has a luminosity similar to
that of Sextans, but unlike Sextans, Leo II sustained a low but
constant star formation rate, leading to an important intermediate age
stellar population \citep{mighell06,koch07}. The reason for a
continuous (low) star formation in Leo II's model is found in the
slightly smaller initial total mass for the same star formation
parameter as Sextans; it prevents the gas outflow.  Carina and Leo II
have about the same mean metallicities.  They probably have probably
experienced a very similar evolution, as indeed seen in
Table~\ref{table3} for their mean properties.  However, the two
galaxies indeed differ: i) by the periods of quiescence which are
longer in Carina, and ii) by the extent of the star formation peaks,
which are between 2 to 3 times longer in Leo II, at intermediate ages.
Model \#587 produced a higher final luminosity than observed, due to
the presence of residual gas and therefore recent star formation, as
in all the models. It is important to recall that we do not aim at
reproducing Leo II properties in fine details, but instead check
whether its main features exist in the series of models we ran.  In
this respect, model \#587 demonstrates that a low level of rather
continuous star formation is indeed achievable.

The extension of the sample of stars with spectroscopic data allowing
measurements of abundance ratios would greatly help the precise
identification of Leo II model.  This is particularly true at [Fe/H]
$> -1.6$ which would confirm or not the apparent plateau in [Mg/Fe].
Even more decisive and more general, one needs to observationally
confirm or disprove the dispersion in [Mg/Fe] at low and moderate
metallicities in low mass dSphs, as it signals the discrete nature of
star formation as advocated by contemporary models.


Table~\ref{table2} gives the fraction of very-metal poor ([Fe/H]$<-3$)
stellar particles in the generic models.  The general trend of our
series of \numsim~ models is that the more extended the star formation
period, the smaller this fraction.  This is clearly illustrated by the
models of Carina and Fornax, as compared to those of Sextans and
Fornax.  Although the tail of very metal-poor stars is small
(from$\sim$ 1 to 6\%), it is still larger than suggested by the
observations to date \citep{helmi06}.  In order to properly address
this particular issue, one needs i) to investigate the impact of more
sophisticated IMFs, such as the one of \citet{kroupa93}, for which the
number of low-mass stars is smaller than for a Salpeter IMF, and ii)
to introduce the peculiar features of star formation at zero
metallicity.  The conditions of transition between population III and
population II star formation depends on the galaxy halo masses, and so
is the galaxy metallicity floor \citep{wise08}.  This opens a
independent new field of investigation.  For the time being, we check
our model predictions against population II properties. Interestingly,
the outflows generated by these very first generation of stars offers
an alternative solution to get rid of the final excess of gas.

\subsection{Mass profiles and velocity dispersions}

\begin{figure}
\resizebox{\hsize}{!}{\includegraphics[angle=0]{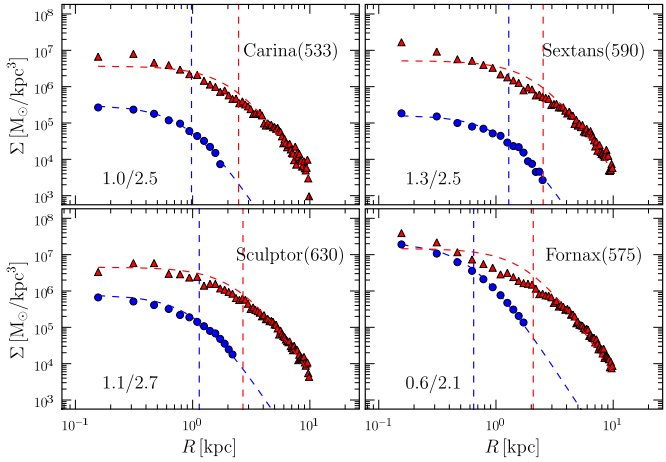}}

\caption{{\rm The stars (blue circles) and 
dark halo (red triangles) density profiles for the four
generic models. Each profile is fitted
by a Plummer model. The positions of the core radii are indicated
by the dashed lines. Their values are given in the left bottom corner 
of each panel.}}
\label{profiles}
\end{figure}

Although our primary goal is not  to match the morphology of the
dSphs, we have considered the shape of the final modeled dSph profiles
within  $R_{\rm{L}}$.  Fig.~\ref{profiles}  displays the  mass density
profiles of the stellar and dark components for our four main targets.
Each model is successfully fitted by a Plummer model, revealing a flat
inner profile, in agreement  with the observations.  The ratio between
the stellar and dark core radii $r_{c}^{\rm{stars}}/r_{c}^{\rm{halo}}$
ranges between $0.2$ and $0.5$,   indicating a spatial segregation  of
the stars relative to the dark halo, as discussed in
\citep{penarrubia08}. Additionally, the line-of-sight stellar velocity
dispersion profiles are  flat in the  enclosed within  the core radii,
with  $6$ to $10\,\rm{km/s}$, corresponding to the  least  and the most massive
dSph, respectively,    hence     well within  the    observed    range
\citep{walker06,munoz06,battaglia08,walker06b}.

\section{Discussion}\label{discussion}

We have shown that the total initial mass $M_{\rm{i}}$ and $c_{\star}$
are the two driving parameters for the evolution of dSphs.  Our models
which describe galaxies in isolation, i.e., not yet taking into
account interactions or accretions that inevitably occur in a
$\Lambda$CDM Universe, already lead to the observed variety of star
formation histories. Decreasing the initial total mass of the galaxies
separates the periods of star formation.

\begin{figure}
\resizebox{\hsize}{!}{\includegraphics[angle=0]{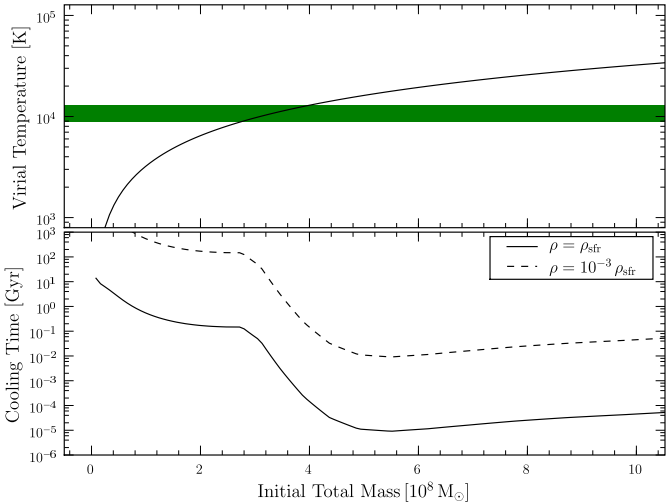}}
\caption{\emph{Top:} The Virial temperature as a function of the total
  initial mass computed from Eq.~\ref{T_gas}, with
  $r_{\rm{c}}=1\,\rm{kpc}$. The lower border of the green band gives
  the Virial temperature of gas with density equal to
  $\rho_{\rm{sfr}}$.  The upper border indicates the Virial
  temperature of gas with density equal to $10^{-3}\,\rho_{\rm{sfr}}$.
  The cooling time is of the order of $0.1\,\rm{Gyr}$. \emph{Bottom:}
  Cooling time of the gas as a function of the total mass, assuming a
  Virial temperature given by Eq.~\ref{T_gas}.  The case of a gas
  density equal to $\rho_{\rm{sfr}}$ is shown by the solid line, while
  the density of $10^{-3}\,\rho_{\rm{sfr}}$ is shown by the dashed
  line. }
\label{virial_temperature}
\end{figure}

The marked differences found between high- and low mass systems is a
key point in understanding the formation of dSphs.  In order to
illustrate our findings, we now focus on cooling time and gas
temperature.  The top panel of Fig.~\ref{virial_temperature} displays
the galaxy's virial temperature as a function of its initial mass
(Eq.~\ref{T_gas}).

The green horizontal band indicates the temperature at which the
cooling time of the gas is equal to $100\,\rm{Myr}$, for two threshold
densities, $\rho_{\rm{sfr}}$ and $10^{-3}\,\rho_{\rm{sfr}}$ that are
representative of the range in gas density observed in the course of
our simulations.  The value of $100\,\rm{Myr}$ is chosen to be short
compared to the galaxy lifetimes.  The lower border of the green band
corresponds to $\rho_{\rm{sfr}}$ while the upper ones correspond to
$10^{-3}\,\rho_{\rm{sfr}}$.  The intersection between the green band
and the curves corresponds to a total mass ranging between $3$ and
$4\times 10^8\,\rm{M_\odot}$, delimiting two regimes distinguished by
long and short cooling times, as can be measured in the bottom panel
of Fig.~\ref{virial_temperature}.  The cooling times of
$\rho_{\rm{sfr}}$ and $10^{-3}\,\rho_{\rm{sfr}}$ as a function of the
galaxy initial masses are shown in solid and dashed lines,
respectively, in the bottom panel of the figure. They are calculated
by taking the virial temperature of the corresponding galaxy mass.
The gas contained in galaxies of total masses larger than $4\times
10^8\,\rm{M_\odot}$, with corresponding virial temperatures larger
than $10^4\,\rm{K}$ is characterized by a short cooling time
($\tau<$100\,\rm{Myr}) at all densities.  This short cooling time is
due to the strong radiative cooling induced by the recombination of
hydrogen above $10^4\,\rm{K}$.  A galaxy which is in this cooling
regime will see its gas loosing a huge amount of energy, sinking in
its central regions and inducing star formation.  This is precisely
what happens in the model \#575 where stars are continuously formed,
generating a metal-rich Fornax-like system.  By decreasing the total
mass from $4$ to $3\times 10^8\,\rm{M_\odot}$, the virial temperature
falls below the peak of hydrogen ionization.  As a consequence, the
cooling function drops and the cooling time is increased by nearly
three orders of magnitude.  Below $3\times 10^8\,\rm{M_\odot}$, the
cooling time is long, since the loss of energy by radiative cooling is
strongly diminished.  In this regime, star formation can only occur in
episodic bursts, separated by periods corresponding to the mean
cooling time.  Carina gives a clear example of such a case.  For even
lower mass systems ($M_{\rm{tot}}<10^{8}\rm{M_\odot}$), the cooling
time is longer than $10\,\rm{Gyr}$, and therefore no stars are
expected to form.

From our simulations, a total mass of $\sim 4\times
10^8\,\rm{M_\odot}$ leads to a luminosity of $\sim
3\times\,10^6\,\rm{L_\odot}$.  Not surprisingly, this luminosity
corresponds to the critical one found in
Section~\ref{global_relations}, below which the galaxies have more gas
left than they have formed stars.
The  sharp cutoff    of     the cooling   function   below    $4\times
10^8\,\rm{M_\odot}$ explains very nicely why  a small decrease in mass
induces a large   drop in luminosity,  seen  in  the middle  panel  of
Fig.~\ref{LFeAgevsMi}. A direct consequence is the constancy of galaxy
masses within $R_{\rm{L}}$ ($\sim  1  - 5 \times  10^7\,\rm{M_\odot}$)
over  the  wide  range   of   dSph   luminosities,  as  discussed   in
Section~\ref{global_relations}  (Fig.~\ref{MvsL}) and   suggested   by
\citet{mateo98}, \citet{gilmore07} and \citet{strigari08}.

The next step  in simulating the formation  and evolution of dSph
galaxies is to study the impact of external processes resulting from a
$\Lambda$CDM    complex environment, like   tidal  mass stripping, ram
stripping, dark matter and gas  accretion, as well as reionization and
self-shielding.  \citet{penarrubia08}  show  that tidal  mass stripping
strongly depends on the spatial  distribution of stars relative to the
dark  halo extent.   If  stars are very  concentrated  in large halos,
dSphs will  be more resilient to  tidal  disruption. Dwarf spheroidal
galaxies may  need to loose nearly  $90\%$ of  their mass before their
star formation and therefore chemical evolution start to be affected.
\citet{mayer06} tackle all the above  listed questions and show how
their mechanisms can  be interlaced.  As  an example, in their  models
gravitational tides help   ram pressure stripping by   diminishing the
overall potential of the dwarf, but  tides induce bar formation making
subsequent   stripping more   difficult.    Reionization prevent  star
formation,  but  self-shielding helps  cooling. Clearly, understanding
the   impact  of these  physical  processes  on  our specific modeling
requires dedicated simulations.





\section{Summary}\label{summary}

We have investigated the formation and evolution of dSph galaxies,
which form a specific class among dwarf systems. Indeed, they reach
the highest metallicities at fixed luminosity and are devoid of gas.

We performed \numsim\, self-consistent Nbody/Tree-SPH simulations of
systems initially consisting of dark matter and primordial gas.  We
have only considered galaxies in isolation.  This has enabled us to
identify the dominant physical ingredients at the origin of the
observed variety in dSph properties.  It has also allowed us to
distinguish which of those are due to the galaxy intrinsic evolution
and for which interactions are required.  The diversity of star
formation histories of the Milky Way dSphs, Carina, Leo II, Sextans,
Sculptor and Fornax have been successfully reproduced in a single
formation scheme.

$\bullet$ The crucial parameter driving the dSph evolution is the
total initial mass (gas + dark matter).  To a smaller extent the star
formation parameter, $c_{\star}$, influences the final galaxy
properties.  In particular, it governs the stellar age distribution by
modifying the time intervals at which star formation occurs.  Since
there is no physical reason for changing $c_{\star}$ from galaxy to
galaxy, we understand its variation, if needed to reproduce the
observations, as an indirect evidence for different interaction
histories.  In a hierarchical galaxy formation scheme, our initial
masses correspond to the halo masses that must be reached along the
merger tree before ignition of the bulk of the star
formation. Besides, the chemical abundance ratios constrain the
timescale of these halo mergers.  For example, we have shown that
Fornax could not be formed without a high initial total mass.  This
could in principle mean an already long accretion history.
Nonetheless, its chemical homogeneity requests that this should be
achieved at a very early epoch, before star formation in independent
smaller halos could have widened its abundance patterns.

$\bullet$ Star formation occurs in series of short periods (a few
hundreds of Myr long).  A high frequency SF mode is analogous to a
continuous star formation, while a low frequency SF mode produces well
separated bursts, easily identified observationally as in the case of
Carina.  The period between the star formation events is governed by
the mass of the systems.  It is caused by the dependence of the gas
cooling time on the galaxy mass: the more massive, the shorter it is.
This mass dependency is explained by the drop of the cooling function
below $10^4\,\rm{K}$.  Systems less massive than $3\times
10^{8}\rm{M_\odot}$ have a virial temperature below $10^4\,\rm{K}$ and
are characterized by weak cooling.  Only episodic periods of star
formation are expected.  On the contrary, massive systems
($M_{\rm{tot}} > 4\times 10^{8}\rm{M_\odot}$) have a virial
temperature above $10^4\,\rm{K}$. They are expected to form stars
continuously and generate more metal-rich dSphs.

$\bullet$ The dSph scaling relations ([Fe/H], M/L) are reproduced and
exhibit very low scatter. We have shown that there is a constant final
total galaxy mass over the wide range of dSph luminosity, which
appears as a direct consequence of the cooling time -mass
relationship.

$\bullet$ The $\alpha$-elements were traced by magnesium.  The
dispersion in the [Mg/Fe] vs [Fe/H] diagrams is negligible for
continuous and/or efficient star formation.  It is enhanced when the
star formation occurs in bursts separated by periods of quiescence, or
when the star formation occurs in peaks of uneven intensities.
Therefore, it is observed in galaxies as different as Sextans and
Carina.  Our study strengthens the need for large and homogeneous
samples of stellar spectra obtained at high resolution, that are
necessary to constrain and improve the models.  They call for large
observing programs dedicated to the chemical signatures at low
metallicities in order to establish the level of homogeneity of the
interstellar medium in the early phases of the galaxy evolution.

$\bullet$ The fraction of very metal-poor stars in our generic models
ranges from $\sim 1$ to 6\%.  Although this is a rather small amount,
it is still larger than suggested by the observations.  In order to
properly address this particular issue, one needs to carefully
investigate the effect of more sophisticated IMFs and the peculiar
features of star formation at zero metallicity.

$\bullet$ Gas is found in all our model galaxies after $14\,\rm{Gyr}$
($\sim 10^7\rm{M_\odot}$).  It clearly points to the need for
interactions that would strip the gas in the course of the dSph
evolution.  The possibility that gas is gradually expelled as halos
merge in a hierarchical formation scenario also deserves careful
consideration.


\begin{acknowledgements}
  We wish to thank the anonymous referee for his comments, which helped in improving the
  contents of this paper.
  Data reduction and galaxy maps have been performed using the parallelized Python
  pNbody package (see http://obswww.unige.ch/~revaz/pNbody/).
  This work was supported by the Swiss National Science Foundation.  
\end{acknowledgements}


\begin{appendix}


\section{Simulation list and parameters}\label{appendix1}

\begin{table*}
    \begin{tabular}{c c c c c c c}
    \hline
    \hline
             N &  $r_{\rm{c}}$  & $f_{\rm{b}}$ & $c_\star$ & $\rho_{\rm{c,tot}}$  & $M_{\rm{tot}}$ & $M_{\rm{gas}}$  \\
              &  $[\rm{kpc}]$       &              &           & $[10^8\,\rm{M_\odot/kpc^3}]$   & $[10^8\,\rm{M_\odot}]$ & $[10^8\,\rm{M_\odot}]$  \\
    \hline
    \hline

    519 & 1.0 & 0.20 & 0.010 & 5.00e-04 & 4.12 & 0.82 \\
    520 & 1.0 & 0.20 & 0.010 & 4.00e-04 & 3.29 & 0.66 \\
    524 & 1.0 & 0.20 & 0.010 & 3.00e-04 & 2.47 & 0.49 \\
    630 & 1.0 & 0.20 & 0.035 & 5.00e-04 & 4.12 & 0.82 \\
    642 & 1.0 & 0.20 & 0.035 & 4.75e-04 & 3.91 & 0.78 \\
    631 & 1.0 & 0.20 & 0.035 & 4.50e-04 & 3.71 & 0.74 \\
    581 & 1.0 & 0.20 & 0.020 & 5.00e-04 & 4.12 & 0.82 \\
    582 & 1.0 & 0.20 & 0.020 & 4.00e-04 & 3.29 & 0.66 \\
    583 & 1.0 & 0.20 & 0.020 & 3.00e-04 & 2.47 & 0.49 \\
    666 & 1.0 & 0.20 & 0.050 & 1.00e-03 & 8.24 & 1.65 \\
    658 & 1.0 & 0.20 & 0.050 & 8.00e-04 & 6.59 & 1.32 \\
    662 & 1.0 & 0.20 & 0.050 & 6.60e-04 & 5.48 & 1.10 \\
    405 & 1.0 & 0.20 & 0.050 & 5.00e-04 & 4.12 & 0.82 \\
    409 & 1.0 & 0.20 & 0.050 & 4.00e-04 & 3.29 & 0.66 \\
    563 & 1.0 & 0.20 & 0.050 & 3.75e-04 & 3.09 & 0.62 \\
    555 & 1.0 & 0.20 & 0.050 & 3.50e-04 & 2.88 & 0.58 \\
    454 & 1.0 & 0.20 & 0.050 & 3.00e-04 & 2.47 & 0.49 \\
    624 & 1.0 & 0.20 & 0.065 & 5.00e-04 & 4.12 & 0.82 \\
    620 & 1.0 & 0.20 & 0.075 & 5.00e-04 & 4.12 & 0.82 \\
    667 & 1.0 & 0.20 & 0.100 & 1.00e-03 & 8.24 & 1.65 \\
    659 & 1.0 & 0.20 & 0.100 & 8.00e-04 & 6.59 & 1.32 \\
    663 & 1.0 & 0.20 & 0.100 & 6.60e-04 & 5.48 & 1.10 \\
    571 & 1.0 & 0.20 & 0.100 & 6.50e-04 & 5.35 & 1.07 \\
    570 & 1.0 & 0.20 & 0.100 & 6.00e-04 & 4.94 & 0.99 \\
    565 & 1.0 & 0.20 & 0.100 & 5.50e-04 & 4.53 & 0.91 \\
    521 & 1.0 & 0.20 & 0.100 & 5.00e-04 & 4.12 & 0.82 \\
    622 & 1.0 & 0.20 & 0.100 & 4.75e-04 & 3.91 & 0.78 \\
    618 & 1.0 & 0.20 & 0.100 & 4.50e-04 & 3.71 & 0.74 \\
    522 & 1.0 & 0.20 & 0.100 & 4.00e-04 & 3.29 & 0.66 \\
    564 & 1.0 & 0.20 & 0.100 & 3.75e-04 & 3.09 & 0.62 \\
    554 & 1.0 & 0.20 & 0.100 & 3.50e-04 & 2.88 & 0.58 \\
    523 & 1.0 & 0.20 & 0.100 & 3.00e-04 & 2.47 & 0.49 \\
    668 & 1.0 & 0.20 & 0.200 & 1.00e-03 & 8.24 & 1.65 \\
    660 & 1.0 & 0.20 & 0.200 & 8.00e-04 & 6.59 & 1.32 \\
    664 & 1.0 & 0.20 & 0.200 & 6.60e-04 & 5.48 & 1.10 \\
    562 & 1.0 & 0.20 & 0.200 & 5.00e-04 & 4.12 & 0.82 \\
    559 & 1.0 & 0.20 & 0.200 & 4.00e-04 & 3.29 & 0.66 \\
    578 & 1.0 & 0.20 & 0.200 & 3.75e-04 & 3.09 & 0.62 \\
    573 & 1.0 & 0.20 & 0.200 & 3.50e-04 & 2.88 & 0.58 \\
    572 & 1.0 & 0.20 & 0.200 & 3.00e-04 & 2.47 & 0.49 \\
    669 & 1.0 & 0.20 & 0.300 & 1.00e-03 & 8.24 & 1.65 \\
    661 & 1.0 & 0.20 & 0.300 & 8.00e-04 & 6.59 & 1.32 \\
    665 & 1.0 & 0.20 & 0.300 & 6.60e-04 & 5.48 & 1.10 \\
    632 & 1.0 & 0.20 & 0.300 & 5.00e-04 & 4.12 & 0.82 \\
    699 & 1.0 & 0.20 & 0.300 & 3.00e-04 & 2.47 & 0.49 \\
    623 & 1.0 & 0.15 & 0.010 & 6.60e-04 & 5.48 & 0.82 \\
    621 & 1.0 & 0.15 & 0.025 & 5.30e-04 & 4.39 & 0.66 \\
    649 & 1.0 & 0.15 & 0.050 & 1.00e-03 & 8.24 & 0.82 \\
    647 & 1.0 & 0.15 & 0.050 & 8.00e-04 & 5.60 & 0.99 \\
    585 & 1.0 & 0.15 & 0.050 & 6.60e-04 & 5.48 & 0.82 \\
    586 & 1.0 & 0.15 & 0.050 & 5.30e-04 & 4.39 & 0.66 \\
    641 & 1.0 & 0.15 & 0.050 & 4.65e-04 & 3.83 & 0.57 \\
    587 & 1.0 & 0.15 & 0.050 & 4.00e-04 & 3.29 & 0.49 \\
    648 & 1.0 & 0.15 & 0.050 & 2.60e-04 & 2.19 & 0.33 \\
    650 & 1.0 & 0.15 & 0.100 & 1.00e-03 & 8.24 & 0.82 \\
    569 & 1.0 & 0.15 & 0.100 & 8.00e-04 & 5.60 & 0.99 \\
    566 & 1.0 & 0.15 & 0.100 & 6.60e-04 & 5.48 & 0.82 \\
    567 & 1.0 & 0.15 & 0.100 & 5.30e-04 & 4.39 & 0.66 \\
    568 & 1.0 & 0.15 & 0.100 & 4.00e-04 & 3.29 & 0.49 \\
    579 & 1.0 & 0.15 & 0.100 & 2.60e-04 & 2.19 & 0.33 \\
    651 & 1.0 & 0.15 & 0.200 & 1.00e-03 & 8.24 & 0.82 \\
    574 & 1.0 & 0.15 & 0.200 & 8.00e-04 & 5.60 & 0.99 \\
    575 & 1.0 & 0.15 & 0.200 & 6.60e-04 & 5.48 & 0.82 \\
    576 & 1.0 & 0.15 & 0.200 & 5.30e-04 & 4.39 & 0.66 \\
    584 & 1.0 & 0.15 & 0.200 & 4.50e-04 & 3.71 & 0.56 \\
    577 & 1.0 & 0.15 & 0.200 & 4.00e-04 & 3.29 & 0.49 \\
    580 & 1.0 & 0.15 & 0.200 & 2.60e-04 & 2.19 & 0.33 \\
    652 & 1.0 & 0.15 & 0.300 & 1.00e-03 & 8.24 & 1.24 \\
    646 & 1.0 & 0.15 & 0.300 & 8.00e-04 & 6.59 & 0.99 \\
    633 & 1.0 & 0.15 & 0.300 & 6.60e-04 & 5.48 & 0.82 \\
    \hline
    \end{tabular}
    \caption[]{Model parameters for the complete set of simulations.}
    \label{parameters_1}
\end{table*}

\begin{table*}
    \begin{tabular}{c c c c c c c}
    \hline
    \hline
             N &  $r_{\rm{c}}$  & $f_{\rm{b}}$ & $c_\star$ & $\rho_{\rm{c,tot}}$  & $M_{\rm{tot}}$ & $M_{\rm{gas}}$  \\
              &  $[\rm{kpc}]$       &              &           & $[10^8\,\rm{M_\odot/kpc^3}]$   & $[10^8\,\rm{M_\odot}]$ & $[10^8\,\rm{M_\odot}]$  \\
    \hline
    \hline

    634 & 1.0 & 0.15 & 0.300 & 5.30e-04 & 4.39 & 0.66 \\
    644 & 1.0 & 0.15 & 0.300 & 4.00e-04 & 3.29 & 0.49 \\
    645 & 1.0 & 0.15 & 0.300 & 2.60e-04 & 2.19 & 0.33 \\
    596 & 1.0 & 0.10 & 0.050 & 1.00e-03 & 8.24 & 0.82 \\
    597 & 1.0 & 0.10 & 0.050 & 8.00e-04 & 6.59 & 0.66 \\
    598 & 1.0 & 0.10 & 0.050 & 6.00e-04 & 4.94 & 0.49 \\
    605 & 1.0 & 0.10 & 0.050 & 4.00e-04 & 3.29 & 0.33 \\
    599 & 1.0 & 0.10 & 0.100 & 1.00e-03 & 8.24 & 0.82 \\
    600 & 1.0 & 0.10 & 0.100 & 8.00e-04 & 6.59 & 0.66 \\
    619 & 1.0 & 0.10 & 0.100 & 7.00e-04 & 5.76 & 0.58 \\
    601 & 1.0 & 0.10 & 0.100 & 6.00e-04 & 4.94 & 0.49 \\
    606 & 1.0 & 0.10 & 0.100 & 4.00e-04 & 3.29 & 0.33 \\
    602 & 1.0 & 0.10 & 0.200 & 1.00e-03 & 8.24 & 0.82 \\
    603 & 1.0 & 0.10 & 0.200 & 8.00e-04 & 6.59 & 0.66 \\
    604 & 1.0 & 0.10 & 0.200 & 6.00e-04 & 4.94 & 0.49 \\
    607 & 1.0 & 0.10 & 0.200 & 4.00e-04 & 3.29 & 0.33 \\
    670 & 1.0 & 0.10 & 0.300 & 1.00e-03 & 8.24 & 0.82 \\
    635 & 1.0 & 0.10 & 0.300 & 8.00e-04 & 6.59 & 0.66 \\
    636 & 1.0 & 0.10 & 0.300 & 6.00e-04 & 4.94 & 0.49 \\
    692 & 1.0 & 0.10 & 0.300 & 4.00e-04 & 3.29 & 0.33 \\
    528 & 0.5 & 0.20 & 0.010 & 2.00e-03 & 4.55 & 0.91 \\
    529 & 0.5 & 0.20 & 0.010 & 1.60e-03 & 3.64 & 0.73 \\
    530 & 0.5 & 0.20 & 0.010 & 1.20e-03 & 2.73 & 0.55 \\
    676 & 0.5 & 0.20 & 0.050 & 4.00e-03 & 9.11 & 1.82 \\
    673 & 0.5 & 0.20 & 0.050 & 3.20e-03 & 7.28 & 1.46 \\
    672 & 0.5 & 0.20 & 0.050 & 2.50e-03 & 5.69 & 1.14 \\
    525 & 0.5 & 0.20 & 0.050 & 2.00e-03 & 4.55 & 0.91 \\
    626 & 0.5 & 0.20 & 0.050 & 1.80e-03 & 4.10 & 0.82 \\
    526 & 0.5 & 0.20 & 0.050 & 1.60e-03 & 3.64 & 0.73 \\
    527 & 0.5 & 0.20 & 0.050 & 1.20e-03 & 2.73 & 0.55 \\
    677 & 0.5 & 0.20 & 0.100 & 4.00e-03 & 9.11 & 1.82 \\
    671 & 0.5 & 0.20 & 0.100 & 2.50e-03 & 5.69 & 1.14 \\
    531 & 0.5 & 0.20 & 0.100 & 2.00e-03 & 4.55 & 0.91 \\
    532 & 0.5 & 0.20 & 0.100 & 1.60e-03 & 3.64 & 0.73 \\
    533 & 0.5 & 0.20 & 0.100 & 1.20e-03 & 2.73 & 0.55 \\
    627 & 0.5 & 0.20 & 0.200 & 4.00e-03 & 9.11 & 1.82 \\
    628 & 0.5 & 0.20 & 0.200 & 3.00e-03 & 6.83 & 1.37 \\
    629 & 0.5 & 0.20 & 0.200 & 2.50e-03 & 5.69 & 1.14 \\
    608 & 0.5 & 0.20 & 0.200 & 2.00e-03 & 4.55 & 0.91 \\
    609 & 0.5 & 0.20 & 0.200 & 1.60e-03 & 3.64 & 0.73 \\
    610 & 0.5 & 0.20 & 0.200 & 1.20e-03 & 2.73 & 0.55 \\
    678 & 0.5 & 0.20 & 0.300 & 4.00e-03 & 9.11 & 1.82 \\
    675 & 0.5 & 0.20 & 0.300 & 3.20e-03 & 7.28 & 1.46 \\
    640 & 0.5 & 0.20 & 0.300 & 2.50e-03 & 5.69 & 1.14 \\
    637 & 0.5 & 0.20 & 0.300 & 2.00e-03 & 4.55 & 0.91 \\
    691 & 0.5 & 0.20 & 0.300 & 1.60e-03 & 3.64 & 0.73 \\
    690 & 0.5 & 0.20 & 0.300 & 1.20e-03 & 2.73 & 0.55 \\
    625 & 0.5 & 0.15 & 0.010 & 2.13e-03 & 4.85 & 0.73 \\
    643 & 0.5 & 0.15 & 0.025 & 2.13e-03 & 4.85 & 0.73 \\
    684 & 0.5 & 0.15 & 0.050 & 4.00e-03 & 9.11 & 1.37 \\
    680 & 0.5 & 0.15 & 0.050 & 3.20e-03 & 7.28 & 1.09 \\
    588 & 0.5 & 0.15 & 0.050 & 2.66e-03 & 6.06 & 0.91 \\
    589 & 0.5 & 0.15 & 0.050 & 2.13e-03 & 4.85 & 0.73 \\
    590 & 0.5 & 0.15 & 0.050 & 1.60e-03 & 3.64 & 0.55 \\
    591 & 0.5 & 0.15 & 0.050 & 1.20e-03 & 2.73 & 0.41 \\
    685 & 0.5 & 0.15 & 0.100 & 4.00e-03 & 9.11 & 1.37 \\
    681 & 0.5 & 0.15 & 0.100 & 3.20e-03 & 7.28 & 1.09 \\
    592 & 0.5 & 0.15 & 0.100 & 2.66e-03 & 6.06 & 0.91 \\
    593 & 0.5 & 0.15 & 0.100 & 2.13e-03 & 4.85 & 0.73 \\
    594 & 0.5 & 0.15 & 0.100 & 1.60e-03 & 3.64 & 0.55 \\
    595 & 0.5 & 0.15 & 0.100 & 1.20e-03 & 2.73 & 0.41 \\
    686 & 0.5 & 0.15 & 0.200 & 4.00e-03 & 9.11 & 1.37 \\
    682 & 0.5 & 0.15 & 0.200 & 3.20e-03 & 7.28 & 1.09 \\
    611 & 0.5 & 0.15 & 0.200 & 2.66e-03 & 6.06 & 0.91 \\
    612 & 0.5 & 0.15 & 0.200 & 2.13e-03 & 4.85 & 0.73 \\
    613 & 0.5 & 0.15 & 0.200 & 1.60e-03 & 3.64 & 0.55 \\
    693 & 0.5 & 0.15 & 0.200 & 1.20e-03 & 2.73 & 0.41 \\
    687 & 0.5 & 0.15 & 0.300 & 4.00e-03 & 9.11 & 1.37 \\
    683 & 0.5 & 0.15 & 0.300 & 3.20e-03 & 7.28 & 1.09 \\
    679 & 0.5 & 0.15 & 0.300 & 2.66e-03 & 6.06 & 0.91 \\
    \hline
    \end{tabular}
    \caption[]{Tab.~\ref{parameters_1} continued.}
    \label{parameters_2}
\end{table*}

\begin{table*}
    \begin{tabular}{c c c c c c c}
    \hline
    \hline
             N &  $r_{\rm{c}}$  & $f_{\rm{b}}$ & $c_\star$ & $\rho_{\rm{c,tot}}$  & $M_{\rm{tot}}$ & $M_{\rm{gas}}$  \\
              &  $[\rm{kpc}]$       &              &           & $[10^8\,\rm{M_\odot/kpc^3}]$   & $[10^8\,\rm{M_\odot}]$ & $[10^8\,\rm{M_\odot}]$  \\
    \hline
    \hline

    638 & 0.5 & 0.15 & 0.300 & 2.13e-03 & 4.85 & 0.73 \\
    695 & 0.5 & 0.15 & 0.300 & 1.60e-03 & 3.64 & 0.55 \\
    694 & 0.5 & 0.15 & 0.300 & 1.20e-03 & 2.73 & 0.41 \\
    537 & 0.5 & 0.10 & 0.025 & 4.00e-03 & 9.11 & 0.91 \\
    538 & 0.5 & 0.10 & 0.025 & 3.20e-03 & 7.28 & 0.73 \\
    539 & 0.5 & 0.10 & 0.025 & 2.40e-03 & 5.46 & 0.55 \\
    534 & 0.5 & 0.10 & 0.050 & 4.00e-03 & 9.11 & 0.91 \\
    535 & 0.5 & 0.10 & 0.050 & 3.20e-03 & 7.28 & 0.73 \\
    536 & 0.5 & 0.10 & 0.050 & 2.40e-03 & 5.46 & 0.55 \\
    547 & 0.5 & 0.10 & 0.050 & 1.80e-03 & 4.10 & 0.41 \\
    546 & 0.5 & 0.10 & 0.050 & 1.20e-03 & 2.73 & 0.27 \\
    540 & 0.5 & 0.10 & 0.100 & 4.00e-03 & 9.11 & 0.91 \\
    541 & 0.5 & 0.10 & 0.100 & 3.20e-03 & 7.28 & 0.73 \\
    542 & 0.5 & 0.10 & 0.100 & 2.40e-03 & 5.46 & 0.55 \\
    549 & 0.5 & 0.10 & 0.100 & 1.80e-03 & 4.10 & 0.41 \\
    548 & 0.5 & 0.10 & 0.100 & 1.20e-03 & 2.73 & 0.27 \\
    614 & 0.5 & 0.10 & 0.200 & 4.00e-03 & 9.11 & 0.91 \\
    615 & 0.5 & 0.10 & 0.200 & 3.20e-03 & 7.28 & 0.73 \\
    616 & 0.5 & 0.10 & 0.200 & 2.40e-03 & 5.46 & 0.55 \\
    617 & 0.5 & 0.10 & 0.200 & 1.80e-03 & 4.10 & 0.41 \\
    696 & 0.5 & 0.10 & 0.200 & 1.20e-03 & 2.73 & 0.27 \\
    689 & 0.5 & 0.10 & 0.300 & 4.00e-03 & 9.11 & 0.91 \\
    688 & 0.5 & 0.10 & 0.300 & 3.20e-03 & 7.28 & 0.73 \\
    639 & 0.5 & 0.10 & 0.300 & 2.40e-03 & 5.46 & 0.55 \\
    698 & 0.5 & 0.10 & 0.300 & 1.80e-03 & 4.10 & 0.41 \\
    697 & 0.5 & 0.10 & 0.300 & 1.20e-03 & 2.73 & 0.27 \\
    \hline
    \end{tabular}
    \caption[]{Tab.~\ref{parameters_1} continued.}
    \label{parameters_3}
\end{table*}

\end{appendix}

\bibliographystyle{aa}
\bibliography{bibliography}

\end{document}